\documentclass[10pt,conference]{IEEEtran}

\usepackage[utf8]{inputenc}
\usepackage{url}
\usepackage{colortbl}
\usepackage{balance}
\usepackage{xspace}
\usepackage{graphicx}
\usepackage{verbatim}
\usepackage{bold-extra}
\usepackage{amsmath}
\usepackage{multirow}
\usepackage{listings}
\usepackage{boxedminipage}
\usepackage{soul}
\usepackage{enumitem}
\usepackage[normalem]{ulem}
\setlist{nolistsep,leftmargin=.5cm}
\usepackage{booktabs}
\usepackage{tabularx}
\usepackage{svg}
\usepackage{calc}
\usepackage{float}
\usepackage{multirow}
\usepackage[normalem]{ulem}
\useunder{\uline}{\ul}{}
\usepackage{amsmath}
\usepackage[most]{tcolorbox}
\definecolor{MidnightBlue}{HTML}{006895}
\definecolor{BoxesBlue}{HTML}{DEECFF}
\definecolor{BoxesYellow}{HTML}{FFF2CC}
\definecolor{StateGreen}{HTML}{91C788}
\definecolor{StateRed}{HTML}{FF8080}
\definecolor{ArrowGreen}{HTML}{61B15A}
\definecolor{ArrowViolet}{HTML}{BA94D1}
\usepackage{caption}
\usepackage{microtype} 
\usepackage[noadjust]{cite}
\usepackage{amsmath,amssymb,amsfonts}
\usepackage{algorithmic}
\usepackage{graphicx}
\usepackage{textcomp}
\usepackage{xcolor}
\usepackage[hidelinks]{hyperref}
\usepackage{ifthen}
\usepackage{wrapfig}
\usepackage{subcaption}
\usepackage{cleveref} 
\usepackage{enumitem}
\usepackage{indentfirst}

\newboolean{showcomments}
\setboolean{showcomments}{false}

\newboolean{showboxes}
\setboolean{showboxes}{true}

\usepackage{adjustbox}
\usepackage{totcount}

\ifthenelse{\boolean{showcomments}}

\setboolean{showcomments}{true}

\ifthenelse{\boolean{showcomments}}
{\newcommand{\nb}[2]{
		\fbox{\bfseries\sffamily\scriptsize#1}
		{\sf\small$\blacktriangleright$\textit{#2}$\blacktriangleleft$}
	}
}
{\newcommand{\nb}[2]{}
}

\newcommand\rev[1]{{\color{black}{#1}}}

\newcommand{\ie}{\textit{i.e.},\xspace}
\newcommand{\eg}{\textit{e.g.},\xspace}
\newcommand{\etc}{\textit{etc.}\xspace}
\newcommand{\etal}{\textit{et al.}\xspace}
\newcommand{\aka}{\textit{a.k.a.}\xspace}

\newcommand{\ir}{{\textsc{reproduction}}\xspace}
\newcommand{\ia}{{\textsc{analysis}}\xspace}
\newcommand{\sd}{{\textsc{solution\ design}}\xspace}
\newcommand{\impl}{{\textsc{implementation}}\xspace}
\newcommand{\crv}{{\textsc{code review}}\xspace}
\newcommand{\ver}{{\textsc{verification}}\xspace}

\newcommand{\trs}{{$\rightarrow$}}

\newcommand{\irs}{{\texttt{\textbf{R}}}\xspace}
\newcommand{\ias}{{\texttt{\textbf{A}}}\xspace}
\newcommand{\sds}{{\texttt{\textbf{SD}}}\xspace}
\newcommand{\impls}{{\texttt{\textbf{I}}}\xspace}
\newcommand{\crvs}{{\texttt{\textbf{CR}}}\xspace}
\newcommand{\vers}{{\texttt{\textbf{V}}}\xspace}

\usepackage{tikz}

\definecolor{bug_red}{rgb}{.84,.23,.29}

\definecolor{info-needed-color}{rgb}{1,.8,.12}

\newcounter{findingcounter}
\newtotcounter{totalfindings}
\ifthenelse{\boolean{showboxes}}
{
    \newcommand{\finding}[1]{%
      \refstepcounter{findingcounter}
      \begin{tcolorbox}[boxsep=1pt,left=2pt,right=2pt,top=1pt,bottom=1pt]%
      \small
      \centering
      \textbf{Finding \arabic{findingcounter}:} #1
      \end{tcolorbox}%
      \addtocounter{totalfindings}{1}
    }
    
  }
{
    \newcommand{\finding}[1]{}
}

\ifthenelse{\boolean{showboxes}}
{
	\newcommand{\rqanswer}[1]{%
		\begin{tcolorbox}[enhanced,skin=enhancedmiddle,borderline={1mm}{0mm}{MidnightBlue},boxsep=3pt]
			\small
			#1
		\end{tcolorbox} 
    }
}
{
	\newcommand{\rqanswer}[1]{}
}

\def\BibTeX{{\rm B\kern-.05em{\sc i\kern-.025em b}\kern-.08em
    T\kern-.1667em\lower.7ex\hbox{E}\kern-.125emX}}

\usepackage[firstpage]{draftwatermark}
\SetWatermarkAngle{0}
\SetWatermarkText{\raisebox{26cm}{%
  \hspace{16cm}%
  \href{https://www.acm.org/publications/policies/artifact-review-and-badging-current}{\includegraphics[width=0.20\linewidth]{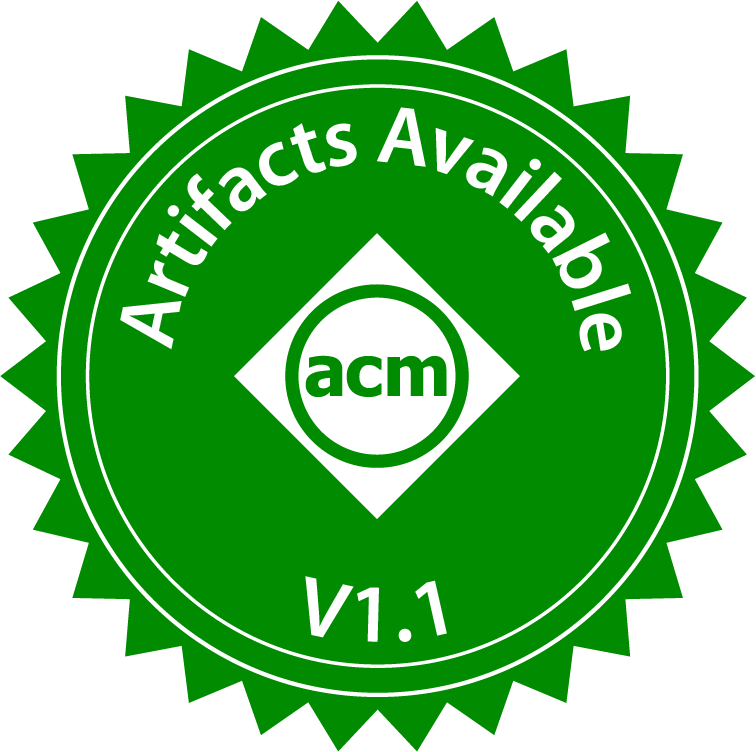}}%
  \href{https://www.acm.org/publications/policies/artifact-review-and-badging-current}{\includegraphics[width=0.20\linewidth]{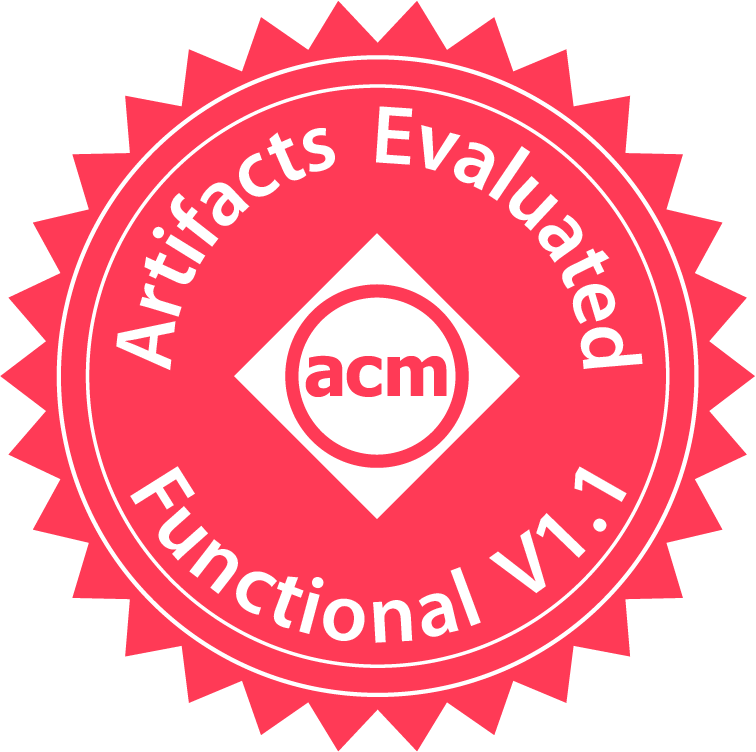}}%
	\href{https://www.acm.org/publications/policies/artifact-review-and-badging-current}{\includegraphics[width=0.20\linewidth]{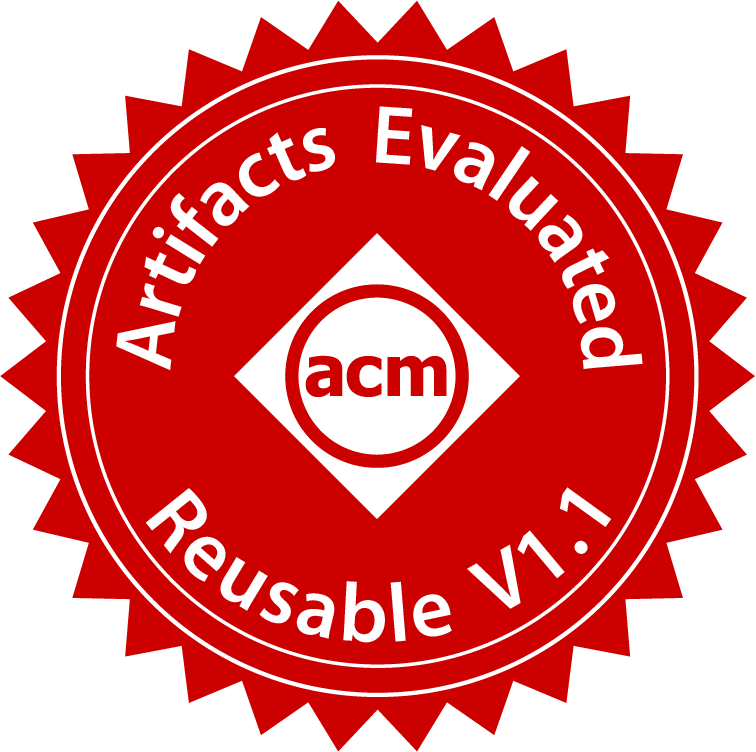}}%
}}

\begin{document}

\title{Decoding the Issue Resolution Process in Practice via Issue Report Analysis: A Case Study of Firefox}

\author{
\IEEEauthorblockN{Antu Saha}
\IEEEauthorblockA{
\textit{William \& Mary}\\
Williamsburg, Virginia, USA \\
\href{mailto:}{asaha02@wm.edu}}
\and
\IEEEauthorblockN{Oscar Chaparro}
\IEEEauthorblockA{
\textit{William \& Mary}\\
Williamsburg, Virginia, USA \\
\href{mailto:}{oscarch@wm.edu}}
}

\maketitle

\thispagestyle{plain}
\pagestyle{plain}

\begin{abstract}
Effectively managing and resolving software issues is critical for maintaining and evolving software systems. Development teams often rely on issue trackers and issue reports to track and manage the work needed during issue resolution, ranging from issue reproduction and analysis to solution design, implementation, verification, and deployment. Despite the issue resolution process being generally known in the software engineering community as a sequential list of activities, it is unknown how developers implement this process in practice and how they discuss it in issue reports. This paper aims to enhance our understanding of the issue resolution process implemented in practice by analyzing the issue reports of Mozilla Firefox. We qualitatively and quantitatively analyzed the discussions found in 356 Firefox issue reports, to identify the sequences of stages that developers go through to address various software problems. We analyzed the sequences to identify the overall resolution process at Firefox and derived a catalog of 47 patterns that represent instances of the process. We analyzed the process and patterns across multiple dimensions, including pattern complexity, issue report types, problem categories, and issue resolution times, resulting in various insights about Mozilla's issue resolution process. We discuss these findings and their implications for different stakeholders on how to better assess and improve the issue resolution process.

\looseness=-1

\end{abstract}

\section{Introduction}
\label{sec:intro}

Issue management is a fundamental process that aims to track and manage the code changes needed to address issues during the maintenance and evolution of a software project. Issue trackers are essential tools that provide the infrastructure to implement issue management~\cite{Zimmermann2009}. Such systems provide a platform for documenting software issues, facilitating discussions among stakeholders, and tracking the work and progress of solving the issues~\cite{Zimmermann2009,Zimmermann2010}. The issue management process assisted by issue trackers, typically involves steps such as issue understanding, triage, replication, and analysis, as well as issue fixing (\aka issue resolution)~\cite{Zimmermann2010,firefox-bug-handling}. 
Issue resolution is a sub-process of issue management that aims to diagnose and resolve the reported problems. 

According to existing literature~\cite{zhang2016literature,saha2015understanding,zeller2009programs,rajlich2011software,eren2023analyzing}, the typical issue resolution process includes steps such as issue reproduction, problem investigation, solution design, solution implementation, and validation/verification, which are sequentially applied to solve issues. However, while this process is meant to be generally applied to any software issue, it is unclear how developers implement it in practice for different problems and contexts and how developers discuss it in issue reports.

\looseness=-1
Understanding the issue resolution process implemented in practice is important for improving software maintenance and evolution processes. By gaining insights into how developers address software problems, we can identify bottlenecks and anomalous process implementations,  align prescribed processes with actual practices, and provide developers with better guidance for issue resolution.  Additionally, studying issue resolution can help identify common patterns and strategies that can be applied to similar problems in the future, and confirm the extent to which the implemented process deviates from the typical, linear resolution process from the literature.

This paper aims to enhance our understanding of the issue resolution process implemented in practice by identifying and analyzing the sequence of steps (\ie stages) that developers perform and discuss in issue reports when solving issues. To that end, we conducted a case study on Mozilla Firefox, a mature and widely-used open-source project. Combining qualitative and quantitative methods, we analyzed the discussions present in a sample of 356 Firefox issue reports to identify the stages of issue resolution that Firefox developers engage in, the sequences of stages that issue discussions form, recurrent transitions between stages present in the sequences, the overall issue resolution process implemented at Firefox, and the recurrent instances of this process to solve a variety of problem types reported in different issue reports. 

Using a multi-coder iterative open-coding methodology, we identified six issue resolution stages (\eg issue reproduction, solution design, implementation, and code review). The stages appear in issue reports with varying frequencies across different issue types (defects, enhancements, and tasks) and problem categories (\eg Crashes, UI Issues, and Code Improvements), and form sequences that represent particular instances of the resolution process at Firefox. The stage sequences reveal frequent relationships among stages, particularly between issue reproduction and analysis; among solution design, implementation, and code review; and among implementation, code review, and solution verification. Additionally, based on analysis of consecutive stages appearing in the sequences (\ie bi-grams), we identified the most common transitions between stages and derived the overall issue resolution process at Firefox from them. Such a process is primarily iterative, deviating from the theoretical linear process found in the literature and Firefox's documentation. In this process, developers go back and forth from one stage to another as needed to solve the issues. Finally, utilizing qualitative analysis of the sequences, we identified 47 issue resolution patterns that represent recurrent instances of the overall process of solving different types of problems. \rev{Two Mozilla developers assessed the usefulness of the patterns, identifying potential use cases to enhance Firefox's resolution process.}
\looseness=-1
	
Our study provides evidence of the iterative and diverse nature of the issue resolution process, which widely deviates from the theoretical linear process from the literature. Our methodology, stage sequences, and patterns serve to identify potential anomalies in the way Firefox developers implement the resolution process.
Our pattern catalog and results can help educate future developers and train newcomers at Firefox in the intricate process of issue resolution. Finally, we advocate for developing advanced tooling to assist developers in recording issue resolution activities more easily, as this can have great benefits for traceability, process assessment, code change rationale management, and more.

In summary, this paper makes the following contributions:
\begin{itemize}
    \item A model of Firefox's issue resolution process implemented in practice, derived from qualitative and quantitative analysis of issue report discussions. 
    \item A novel catalog of 47 patterns of issue resolution, derived from qualitative analysis of issue discussions. The patterns represent instances of Firefox's resolution process, employed by Firefox developers to address different types of software problems.
    \item A comprehensive analysis of the derived process and patterns, across different types of issues and problem categories, showing that Firefox's issue resolution is a diverse and iterative process, which deviates from the prescribed linear process from the literature.
    \item A novel dataset with annotated issues and related artifacts that enables further research in this area. We publicly release this data and the derived catalog, scripts, and other artifacts useful to validate and replicate our study~\cite{repl_pack}.
\end{itemize}

\section{Background, Problem, and Motivation}
\label{sec:background}

\subsection{Issue Management during Software Evolution}

\textbf{Issue management} aims to track and manage all the change requests of a software system, including new feature developments, non-functional implementations, defect corrections, and enhancements to existing functionality~\cite{Zimmermann2010,zeller2009programs,rajlich2011software}. 
Issue trackers, such as Bugzilla~\cite{bugzillaBugzilla}, Jira~\cite{jira}, and GitHub Issues~\cite{github}, are communication and social platforms that allow coordination among different stakeholders around the process of issue management~\cite{bertram2009social,bertram2010communication,xia2013accurate}.
\textbf{Issue reports} are the main artifacts created and used during issue management~\cite{Zimmermann2010}, and  
typically include an issue title, a detailed issue description, metadata  (\eg issue severity and priority, operating platform, and product versions affected by the issue), and attachments (\eg system logs and screenshots). 
\looseness=-1

During the process of managing and solving the issues, stakeholders change the status of the issues (\eg from \textit{Assigned} to \textit{Resolved} and from \textit{Verified} to \textit{Closed}~\cite{eren2023analyzing,bugzila-bug-lifecycle}), and engage in discussions when needed. These discussions are recorded in \textbf{issue comments}, which document relevant information about the reported problems~\cite{zeller2009programs}, including possible circumstances in which the problem occurs, potential causes, how the code should be changed to address the issue, and more.
\looseness=-1

\subsection{Issue Resolution within the Issue Management Process}
\label{sub:background_issue_res}

The literature has defined the different phases that compose the issue management process.  Zhang \etal~\cite{zhang2016literature} report three major phases: issue understanding, triaging, and fixing (\aka \textbf{issue resolution}). In the first step, triagers read and understand the report to determine the issue type, priority, and severity. In the second step, the triagers assign the issue to an expert developer. Finally, in the issue resolution phase, the assigned developer locates the code that needs to be changed in response to the issue and implements the desired code change.

K. Saha \etal~\cite{saha2015understanding} define similar phases, including a fourth phase called issue verification, in which a developer verifies that the code change indeed addresses the issue and conforms with the quality standards.
Zeller~\cite{zeller2009programs} includes issue reporting, duplicate report identification, and fix delivery as part of the issue management process. Zeller also decomposes the issue resolution phase for bug reports into multiple steps: bug reproduction, isolation and localization, and fix implementation.
\looseness=-1

Rajlich~\cite{rajlich2011software} defines \textit{software change} as a general process to modify and evolve software systems, based on change requests.
Rajlich defines seven phases of software change: initiation, concept location, impact analysis,  actualization, refactoring, verification, and conclusion. Initiation includes prioritization of change requests. All these phases except for initiation and conclusion are part of issue resolution, which aims to analyze and implement the solution to an issue.

In this paper, we investigate how developers perform issue resolution, a critical sub-process of issue management that aims to diagnose and solve the reported problems~\cite{eren2023analyzing}. 
This \textbf{issue resolution process} includes reproducing the reported bugs (for bug reports), understanding and analyzing the issues, designing a solution to the issue, implementing the solution, and validating/verifying the quality of the implementation.
\looseness=-1

\subsection{The Issue Resolution Process Implemented in Practice}

Although the issue resolution process includes different activities as discussed above, it is unclear how this process is implemented in practice to address \rev{different kinds of issues}, under various operating circumstances. As we discuss in the related work section (\Cref{sec:related_work}), prior work has focused on studying different aspects of issue reports and their management, including the overall issue management process based on issue status transitions, but no prior work has studied in detail the process that developers implement in practice to analyze and solve issues. We fill in this knowledge gap by analyzing issue discussions and studying  \textbf{patterns of issue resolution}. 
\looseness=-1

We motivate our work by discussing two examples of issue resolution at Mozilla Firefox~\cite{mozilla-firefox}. 
Issue report \#1029919~\cite{firefox-bug-1029919} describes a buggy behavior in the way Firefox renders a web page: when the user hovers over HTML buttons on a page, Firefox draws a border around the button.
The issue report contains rich information in the issue comments that help us understand the process followed to solve this issue. 
At first, developers reported multiple bug reproduction attempts, asking for additional information from the reporter. After it was successfully reproduced, the developer assigned to solve the issue posted the result of his investigation, describing the potential problem location and cause. The developer then attached two fixing patches describing the root cause,  the solution, and the potential impact of the solution on the system. The patches were reviewed by another developer and after the original developer corrected a few problems, the code reviewer inspected and approved the implementation. The code change was then integrated into Mozilla's code base, the reporter verified the fix, and a triager closed the issue.

Another example is issue \#1718748~\cite{firefox-bug-1718748}, which describes a failure in Firefox's cross-platform component %
that handles UI rendering. 
The issue states that some buttons in Firefox's toolbar customization UI become invisible when switching to a dark theme. 
The reporter is a QA member who identified a prior commit and issue that could have introduced the bug (via Mozilla's mozregression tool~\cite{mozregression}). The developer responsible for that prior commit and issue, assigned to solve the issue, \rev{provided a patch with the description of the code change to correct the defect}. Another developer reviewed the code change (via the Phabricator code review tool~\cite{phabricator}) and a QA member then successfully verified the solution, marking the issue as \textit{Verified} and \textit{Fixed}.
\looseness=-1

Both examples illustrate different ways to resolve bugs related to Firefox's web page and UI rendering. 
In the first example, we observe all the expected major steps of issue resolution, however, in the second example, the process did not include any issue reproduction and analysis.
In both cases, these issue resolution steps were performed by different stakeholders, 
recording the activities and the relevant information obtained during the issue resolution. While the nature of the problems might have been different, it is clear that the issue resolution process that we would expect from theory can be implemented and recorded in issue reports in different ways. 
Our goal is to investigate these different approaches and determine if there are recurrent patterns in the process of solving different kinds of issues. We do so by qualitatively analyzing the discussions that developers document in issue reports.  
\looseness=-1

\subsection{The Issue Resolution Process at Mozilla Firefox}
\label{sub:firefox_process}

We selected Mozilla Firefox~\cite{mozilla-firefox} as the subject of our study because (1) it is a mature and widely-used project with 19+ years of evolution, and (2) it has well-documented 
practices for issue management~\cite{firefox-bug-handling} and software development (\eg patching~\cite{firefox-patching}, code quality~\cite{firefox-code-quality,firefox-reviewer-checklist}, testing~\cite{firefox-security-bugs}, and  debugging~\cite{Working-Firefox}), which allow us to understand Firefox's issue resolution process in detail.  Mozilla Firefox is a multi-language, multi-platform open-source project that uses BMO, an adapted version of the Bugzilla issue tracker~\cite{bugzillaBugzilla}, to manage all the changes made to Firefox's source code~\cite{firefox-patching}.
\looseness=-1

\rev{All of Firefox's code changes} are documented in issue reports by end-users, community members, QA members, and developers during system usage, testing, and analysis~\cite{firefox-bug-pipeline}. Firefox has three \textbf{issue report types}~\cite{mozilla-bug-types}: defects, enhancements (\ie user-facing improvements), and tasks (\ie back-end improvements). These issues are triaged differently by a rotating group of engineering managers who are owners of a Firefox component and by QA members~\cite{firefox-triage,firefox-feature-triage,firefox-security-approval}. These members assess the issues and assign a correct issue type, severity, priority, target release, and other metadata (\eg security flags~\cite{firefox-security-approval}) to better prioritize and manage the problems. QA members, component owners, and developers are in charge of determining the resolution state of the issues (\eg \textit{Resolved - Won't fix} or \textit{Verified - Fixed}~\cite{firefox-feature-triage}). 

The open nature of the project makes Firefox's software development, and in particular issue resolution, a worldwide and distributed process. Developers are assigned to issues and work on one or more patches to address the problems. For diagnosing and solving defects, Firefox provides guidelines for using various debugging tools across different platforms~\cite{Working-Firefox}. 
Once the patches are completed, they are attached to the issue reports, requiring a code review through the Phabricator tool~\cite{phabricator}. The tool posts comments on issues whenever a code review is submitted. The code reviewer is mainly a component owner or peer, a newcomer mentor, and/or any other developer familiar with the modified code or module~\cite{firefox-patching}. The patch is tested in Try~\cite{firefox-try}, a system for running automated tests without integrating patches into Firefox's code base. Once the patches are approved, they are integrated (\aka landed), by the code reviewer, into the `autoland' repository, where regression tests are executed~\cite{testing-firefox-ml}. 
Once the tests pass and the code changes are further validated/verified by the QA team, they are merged by `code sheriffs' into `mozilla-central',  Firefox's main development repository~\cite{testing-firefox-ml}. Merging into `mozilla-central' occurs periodically or on demand (\eg when critical security fixes are validated)~\cite{shipping-firefox,testing-firefox-ml}. 

During the resolution process, the status of the reports is updated accordingly (\eg from \textit{Assigned} to \textit{Verified and  Fixed}). 
Information relevant to the issue (\eg failing regression test results), obtained at any moment during the process, may be posted as an issue comment. For example, failing regression test results are posted in the issues. 
Code changes in `mozilla-central' are integrated into `mozilla-beta' for additional quality assurance during a four-week beta cycle. After this, a release candidate build is generated, tested thoroughly, and made available as the next version of Firefox~\cite{shipping-firefox}.

\section{Study Methodology}
\label{sec:methodology}

This study aims to investigate how the issue resolution process is implemented in practice at Mozilla Firefox to solve various software problems and tasks described in issue reports.  We investigate the major stages of the issue resolution process, described in \Cref{sub:background_issue_res}, and how developers\footnote{We hereon use \textit{developers} to refer to all stakeholders involved in issue resolution: programmers, reporters, QA members, \etc}  follow them to solve a variety of problem categories (\eg crashes, UI issues, or refactoring changes) reported in various issue report types (defects, enhancements, or tasks). 
The study addresses the following research questions (RQs):
\looseness=-1

\begin{enumerate}[label=\textbf{RQ$_\arabic*$:}, ref=\textbf{RQ$_\arabic*$}, itemindent=0cm,leftmargin=1cm]
	\item \label{rq:stages}{\textit{What issue resolution stages are found in issue reports?}} 
	\item \label{rq:interactions}{\textit{How do the resolution stages interact with each other?}}
	\item \label{rq:process}{\textit{What is the overall process of issue resolution?}}
	\item \label{rq:patterns}{\textit{What resolution patterns are found in issue reports?}}
	\item \label{rq:pattern_usefulness}{\rev{\textit{What are the potential use cases of the patterns?}} }
\end{enumerate}

\ref{rq:stages} investigates the major stages that Mozilla developers go through to address reported issues and how frequently these stages are discussed in issue reports. \ref{rq:interactions} investigates how these stages interact with one another, including how frequently these stages co-occur in issue reports. \ref{rq:process} investigates the overall issue resolution process at Mozilla Firefox.  \ref{rq:patterns} investigates recurrent instances of the resolution process, expressed as sequences of stages. \rev{\ref{rq:pattern_usefulness} examines the potential applications of the derived patterns for Mozilla developers.}
\looseness=-1

\subsection{Issue Collection}
\label{sub:issue_collection}

Mozilla's BMO is the centralized system for managing the issues of Firefox desktop and mobile~\cite{mozilla-products}. In this study, we focused on the desktop version of Mozilla Firefox, studying the issues of its two main components: \textit{Firefox} and \textit{Core}.  The \textit{Firefox} component (\aka \textit{product} in BMO) implements the graphical user interface (GUI) of the web browser, while the \textit{Core} component includes essential functionality such as web page rendering, web browsing, and networking services.

Our study focused on \textit{FIXED} and \textit{RESOLVED} issue reports
for the selected components. To obtain recent issues within a significant period of system evolution, we downloaded all the issues created from January 1st, 2010 to April 30th, 2023 using Bugzilla's API~\cite{bugzilla-api}, including their title/summary, comments (which contain the issue description), and relevant metadata: creation time, resolution time, and others.
\rev{From 199,271 downloaded issues ($\approx$164.7k/34.5k for Core/Firefox), we randomly sampled 384 issues for analysis. This is a statistically significant sample, at a 95\% confidence level and 5\% error margin, that captures the diversity and characteristics of the entire population of \textit{Core} and \textit{Firefox} issues. This is evidenced by comparing our sample and the entire issue population in terms of the proportion of issue types (defects: 71.1\% vs 70.1\%, enhancements: 16.9\% vs 16.1\%, and tasks: 12\% vs 13.5\%), the proportion of issues per product (Core: 81.5\% vs 82.7\% and Firefox: 18.5\% vs 17.3\%), average \# of comments per issue (13.4 vs 14.6), and average resolution time (81 vs 88 days). 
	} 
The 384 issues contain 13.4 (9) comments, 30.27 (16) paragraphs, and 56.73 (25) sentences on average (median). 
\looseness=-1

\subsection{Issue Annotation}
\label{sub:issue_annotation}

\subsubsection{\textbf{Goals and Overview}}
\rev{
We qualitatively analyzed all the information provided in the issues, annotating textual content related to issue resolution by employing an iterative \textit{open coding} methodology~\cite{spencer2009card}. The annotation process was conducted by six Ph.D. students and one professor (\aka \textit{annotators}), including the authors of this paper.  The annotators have 1-9 years of research experience (particularly in qualitative text analysis), and five of them have 1-4 years of industry experience.}
\looseness=-1

\rev{The annotation targeted all the textual content written by different stakeholders in issue comments and aimed to identify: (1) \textit{themes} or \textit{codes} about different activities performed to resolve the issue 
(\eg reproduction attempts or a code review), and (2) the types of problems described in the issues (\eg crashes, UI issues, \etc), which we call \textbf{problem categories}.}

\rev{\subsubsection{\textbf{Annotation Tool and Unit}} We used the Hypothesis annotation tool~\cite{hypothesis} to directly annotate the web pages of the issue reports. The tool allowed us to collaboratively assign \textit{codes} to text snippets in the issue threads, modify the assigned codes, and discuss the annotations.

We coded \textit{text snippets} in the issue comments. The minimal annotation unit was a complete sentence. Since one or more sentences may convey the same type of
information (\ie a given resolution activity), the annotation included individual sentences, multiple sentences, paragraphs, or even entire comments. A single textual snippet was allowed to be coded with one or more codes.

\subsubsection{\textbf{Code Catalog and Coding Guidelines}}
We maintained a  \textit{code catalog} via a  Google spreadsheet shared among the annotators. The catalog included a list of codes,  code descriptions, rules to apply the codes, and text snippets from annotated issues used as examples. The code catalog also included a list of problem categories, with detailed definitions and examples of annotated issues. We also maintained a shared Google document with detailed guidelines of the annotation procedure, coding rules, and necessary resources for annotating the issues (\eg official Mozilla documentation to get familiar with Firefox's resolution process and a glossary of annotation terminology). Both the catalog and  guidelines were built from scratch and developed by all the annotators incrementally and collaboratively.}
\looseness=-1

\rev{\subsubsection{\textbf{Annotation Procedure}} 
We adopted an iterative multi-coder open-coding methodology wherein each issue report was annotated and validated by at least two annotators.  The 384 issues were distributed evenly among the seven annotators, who iteratively examined, annotated, and validated the issue comments in batches of 30-50 issues.  
The first annotator assigned codes to text snippets in the comments, and a second annotator reviewed these annotations for accuracy and completeness. Discrepancies were resolved in reconciliation sessions. Annotator roles alternated across batches, with each person either annotating from scratch or reviewing the annotations by the first annotator. To avoid fatigue and reduce potential mistakes, the annotators annotated small sets of issues with breaks in between.
\looseness=-1

The overall process for a single issue involved the first annotator thoroughly reviewing the issue, including attached patches, linked commits, and metadata (\eg\ issue commentators, tags, and status), to identify/annotate relevant content and the problem category. Codes were assigned based on the content's meaning and the code catalog. The second annotator then reviewed these annotations, verifying their accuracy, suggesting additional codes, or flagging mistakes. After processing a batch, both annotators discussed disagreements to reach a consensus.

To establish the initial coding framework, two researchers annotated the first batch of 30 issues, creating an initial set of codes and problem categories. These were refined through discussion sessions, resulting in complete definitions, examples, and rules for applying the codes. This initial annotation informed the creation of the coding guidelines, which included resources for understanding issues and general annotation rules.

Before annotating the remaining issues, training sessions were conducted with the other annotators to review the coding guidelines, discuss examples from the initial batch, and solve misunderstandings. Throughout the entire annotation process, the code catalog was continuously updated, with changes such as new codes, code merges, or renames collectively agreed upon and promptly communicated. When the catalog was updated, previously coded issues were revisited to ensure consistency. Regular communication via Zoom meetings and Slack discussions was essential to maintain the accuracy and uniformity of the catalog and annotated content.

}

\rev{\subsubsection{\textbf{Annotation Results and Inter-coder Agreement}} During the annotation process, 28 issues, that were pull requests (PR) automatically created by the issue tracker, were discarded.
In summary, we annotated \textbf{3,707 textual snippets} in 2,574 issue comments across 356 issue reports. The annotation process resulted in 
\textbf{22 issue resolution codes}, and \textbf{17 problem categories} which we further grouped into \textbf{3 problem classes}. \Cref{tab:stages,tab:problem_categories} show examples of these elements;
our replication package contains the full catalog of codes and problems~\cite{repl_pack}. 

The annotators agreed on 3,438 annotations with an agreement rate of $\approx$93\% and a Cohen's kappa of 0.92, which indicates a high overall agreement~\cite{Cohen}. Common
sources of disagreement (269/3,707 text snippets) included misunderstandings due to ambiguous comments or unclear code definitions. If both annotators were unable to reach an agreement, a third annotator reviewed the issue to resolve the conflict. 
\looseness=-1
}

\subsection{Inferring and Analyzing Issue Resolution Stages}
\label{sub:resolution_stages}

The 22 codes obtained from the issue report annotation represent the information about activities performed by developers during issue resolution.
We implemented two steps for inferring the issue resolution stages from the annotation codes. In the first step, we qualitatively analyzed the code catalog and annotated issues and identified 13 codes (\ie actionable codes) that signified specific actions performed to directly address the problems (\eg reproducing the problem or implementing a solution as a code change). In the second step, we engaged in an analysis of issues/codes and a discussion to categorize the 13 codes for inferring \textit{issue resolution stages}. 
\looseness=-1

\begin{table}[t]
\caption{Issue Resolution Stages}

\label{tab:stages}
\resizebox{\columnwidth}{!}{%
\begin{tabular}{c|c|c|c}
\hline
\textbf{Stage}                & \textbf{Description}                                                                                                                                             & \textbf{Annotation Codes}                                                                                        & \textbf{\# of Issues} \\ \hline
\textbf{\ir (\texttt{\textbf{R}})}      & \begin{tabular}[c]{@{}c@{}}Developers attempt to\\ reproduce the issue.\end{tabular}                                                                             & \sc{REP\_ATT}                                                                                                        & 47 (13.2\%)        \\ \hline
\textbf{\ia (\texttt{\textbf{A}})}          & \begin{tabular}[c]{@{}c@{}}Developers analyze the issue\\ by reviewing the problem,\\ identifying the problem cause,\\ or locating the relevant code.
	\\
 \end{tabular}   & \begin{tabular}[c]{@{}c@{}}\sc{PROB\_LOC},\\ \sc{PROB\_REV},\\ \sc{CAUS\_IDENT}\end{tabular}                               & 134 (37.6\%)       \\ \hline
\textbf{\sd (\texttt{\textbf{SD}})} & \begin{tabular}[c]{@{}c@{}}Developers discuss how to\\solve the issue, \ie propose\\a potential solution or\\review a proposed solution.\end{tabular} & \begin{tabular}[c]{@{}c@{}}\sc{POT\_SOL\_DES},\\ \sc{SOL\_REV}\end{tabular}                                                & 150 (42.1\%)       \\ \hline
\textbf{\impl (\texttt{\textbf{I}})}    & \begin{tabular}[c]{@{}c@{}}Developers make the\\ necessary code changes\\ to resolve the issue.\end{tabular}                                           & \sc{CODE\_IMPL}                                                                                                       & 328 (92.1\%)       \\ \hline
\textbf{\crv (\texttt{\textbf{CR}})}     & \begin{tabular}[c]{@{}c@{}}Developers review the\\ implemented code changes.\end{tabular}                                                                                & \sc{CODE\_REV}                                                                                                        & 264 (74.2\%)       \\ \hline
\textbf{\ver (\texttt{\textbf{V}})}      & \begin{tabular}[c]{@{}c@{}}Developers verify the\\solution by testing the\\ implemented code changes.\end{tabular}                                           & \begin{tabular}[c]{@{}c@{}}\sc{SOL\_VER},\\ \sc{UPLIFT\_APRV},\\ \sc{IMPL\_REV},\\ \sc{COL\_PROB\_ANA},\\ \sc{COL\_POT\_SOL}\end{tabular} & 146 (41\%)         \\ \hline
\end{tabular}%
}
\end{table}

\begin{table}[]
\caption{Problem Categories and Classes}
\label{tab:problem_categories}
\resizebox{\columnwidth}{!}{%
\begin{tabular}{c|c|c|c}
\hline
\textbf{Problem Class}  & \textbf{Problem Categories (Examples)}        & \textbf{\# of Categ.} & \textbf{\# of Issues} \\ \hline
{Implementation} & UI Issue, Feature Development, Crash       & 12                 & 261                \\ \hline
{Refactoring}    & Code Improvement, Unnecessary Code Removal & 2                  & 51                 \\ \hline
{Testing}        & Test Failure, Test Update, Flaky Tests     & 3                  & 44                 \\ \hline
\end{tabular}%
}
\end{table}

The first step was necessary because 9 of the 22 codes were either: (1) requests to perform an action, not an action in itself; or (2) cross-cutting actions, which can be performed at any stage of the resolution process. {\sc{solution\_review\_request}} \rev{is an example of a request}, which represents a petition, made by a developer to another one, to review a proposed solution to the problem. 
{\sc{solved\_by\_other\_issue}} is an example of a cross-cutting code that represents an issue resolved in another issue. 
\looseness=-1

Based on the qualitative analysis, we identified six different issue resolution stages, namely:
\ir (\texttt{\textbf{R}}), \ia (\texttt{\textbf{A}}), \sd (\texttt{\textbf{SD}}), \impl(\texttt{\textbf{I}}), \crv (\texttt{\textbf{CR}}), and \ver (\texttt{\textbf{V}}). Each stage is represented by one to five actionable codes, each code belonging to a single stage. Examples of codes for the \ia stage (\texttt{\textbf{A}}) are {\sc{problem\_localization}} and 
{\sc{cause\_identification}}.
\Cref{tab:stages} shows all the stages with their description and codes.

To answer \ref{rq:stages}, we analyze the frequency in which the six stages appear in the issue reports, across different report types and problem classes and categories.

\subsection{Analysis of Stages Sequences and Process Inference}
\label{sub:process}

When the identified stages are aggregated in the order in which codes appear in the issue report (\ie chronologically), they create a sequence of codes, which we can then examine to understand the process adopted to resolve the issue. For example, issue \#1363344's~\cite{firefox-bug} annotation code sequence is: 
{\sc{code\_implementation}},  {\sc{code\_review}},  {\sc{code\_review}}, {\sc{code\_review}}. We created a stage sequence by utilizing the code sequence and the code-stage mapping for each issue. 
For example, for the above code sequence of issue \#1363344~\cite{firefox-bug}, the derived stage sequence is: \texttt{\textbf{I,CR,CR,CR}} which we simplified as \texttt{\textbf{I,CR}} by merging consecutive repeating stages. This process was applied to all the issues.

To answer \ref{rq:interactions}, we counted the bi-grams and tri-grams appearing in the stage sequences, as well as the number of issues where these n-grams appear. Bi-grams are pairs of consecutive stages, while tri-grams are triplets of consecutive stages in the sequences. We also analyzed the frequency with which the stages appear at the beginning or end of the sequences.
\looseness=-1

To answer \ref{rq:process}, we constructed a graph representing the overall issue resolution process, where the nodes correspond to the stages and the edges represent the transitions between stages. This graph was constructed based on the most frequent bi-grams found in the sequences and serves to validate the patterns of issue resolution we derive as part of \ref{rq:patterns} (see \Cref{sub:resolution_patterns}). 
\looseness=-1

\subsection{Inferring Issue Resolution Patterns}
\label{sub:resolution_patterns}
To answer \ref{rq:patterns}, we engaged in a qualitative analysis of the stage sequences and derived issue resolution patterns by grouping similar stage sequences into coarse-grained sequences. The derived patterns correspond to instances of the derived issue resolution process in \ref{rq:process}.

\subsubsection{\textbf{Pattern Notations}} To communicate the issue resolution patterns clearly and analyze them in different dimensions, we represent the patterns as a string based on  three  notations: 
\begin{itemize}
    \item \textbf{A?}  indicates that stage A is optional;
    \item \textbf{(A$\mid$B)} indicates that either A or B or both stages appear; 
    \item \textbf{(A,B,...,Z)+} indicates that stages A, B, ..., and Z appear more than once, and at least one subsequence of two or more stages (A,B or B,Z or A,B,Z, \etc) appears more than once.
    \looseness=-1
\end{itemize}

\subsubsection{\textbf{Deriving Issue Resolution Patterns}} At first, we identified the stage sequences where the 3rd notation, (A,B,...,Z)+, is applicable and created issue resolution patterns for those stage sequences applying the notation. For example, issue \#991812~\cite{firefox-bug-991812} with the stage sequence \texttt{\textbf{I,CR,I,CR,I,CR,V,I,V}} has \texttt{\textbf{I}}, \texttt{\textbf{CR}}, and \texttt{\textbf{V}} appearing more than once and the sub-sequence \texttt{\textbf{I,CR}} appears more than once. Hence, the sequence can be collapsed to create the issue resolution pattern \texttt{\textbf{(I,CR,V)+}}. With this notation, the order of the stages does not matter.

Second, we created groups of stage sequences that differ only by one or two stages in order and qualitatively analyzed each sequence to understand the differences among the sequences. We aimed to represent the sequences using the first two notations (\ie A? and (A$\mid$B)) to form a coarse-grained sequence. For example, issues \#698552~\cite{firefox-bug-698552}, \#676248~\cite{firefox-bug-676248}, and \#730907~\cite{firefox-bug-730907} have the stage sequences ``\texttt{\textbf{SD,I,CR}}", ``\texttt{\textbf{SD,I,CR,I}}'', and ``\texttt{\textbf{SD,I,CR,V}}", respectively. Here, all three stages, \sds, \impls, and \crvs, are included in the three issues. However, the sequences only differ by the last stage: \impls or \vers is present for the last two issues while it is not present in the first one. Hence, we can create a common pattern for these three issues, \ie\ \texttt{\textbf{SD,I,CR,(I$\mid$V)?}} which will represent all three sequences.

\rev{We meticulously created this grouping} by considering several factors (\eg the \# of issues per sequence, the presence of unique stages per sequence, and the issue resolution process of each issue in the group) so that we would not lose information or create any misleading sequence that does not represent the actual resolution process. For example, we could create a group for the sequences \texttt{\textbf{I}} and \texttt{\textbf{I,CR}} by making \texttt{\textbf{CR}} as an optional stage (\ie  \texttt{\textbf{I,CR?}}). However, the first sequence is found for 21 issues and the second sequence is found for 50 issues which implies these two sequences are already widely used and can represent two distinct ways of issue resolution. In the first sequence, no \texttt{\textbf{CR}} is performed, whereas in the second, it is performed to resolve the issue. Hence, we did not create a group from these two sequences.
\looseness=-1

In all qualitative steps, one researcher qualitatively analyzed the issue and made necessary changes by documenting the rationale behind each change which was reviewed and validated by the second researcher. Both researchers continuously discussed the patterns and solved any disagreements.

\rev{\subsubsection{\textbf{Pattern Derivation Results and Pattern Categorization}} Our analysis resulted in 47 distinct issue resolution patterns -- the  10 most frequent patterns are shown in \Cref{tab:patterns}. The patterns contain 1-6 stages and appear in 1-64 issues (7.6 issues on average). 
The more unique stages and the more interacting stages a pattern has, the more complicated a pattern is. We argue that the complexity of a pattern reflects the effort developers invest in resolving an issue, which can be quantified by the number of stages in the sequences associated with the pattern. Therefore,  we categorized the patterns as \textit{simple} or \textit{complex} based on the average number of stages in their sequences. Since the distribution of these averages is not skewed (see our replication package for the distribution~\cite{repl_pack}), the mean serves as a threshold for classification. Specifically, the process involves calculating the average number of stages ($P_a$) for each pattern, determining the overall mean across the patterns ($T=6.2$ stages), and classifying a pattern as complex if $P_a > T$ or simple if $P_a \leq T$. In \Cref{sub:results_patterns}, we discuss the pattern catalog and compare it with the derived process from \ref{rq:process} to answer \ref{rq:patterns}.}
\looseness=-1

\rev{
\subsection{Investigating Potential Use Cases of the Derived Patterns}

To answer \ref{rq:pattern_usefulness}, we conducted semi-structured interviews with two Mozilla developers, aimed to gather detailed feedback from them on the usefulness of the resolution patterns. The interviews were conducted over Zoom for 60 minutes and were structured into four sections: 

\begin{enumerate}
	\item \underline{Participant's Background}: Participants were asked to share their background and experience in software development and issue resolution at Mozilla and other companies.
	
	\item \underline{Mozilla's Issue Resolution Process}: Participants were asked to describe Mozilla's issue resolution process (both prescribed by Mozilla and implemented by developers) as well as the specific approaches they follow.
	
	\item \underline{Research Presentation}: The research team presented the study's goals, methodology, and findings, including the identified patterns. Participants were encouraged to ask questions about the patterns and findings.
	
	\item \underline{Question-Answering}: Participants were asked 11 questions that prompted for feedback on the identified patterns, with a focus on understanding their potential benefits for Mozilla.
\end{enumerate}

Follow-up questions were asked when additional information was needed. The interview questionnaire, protocol, and anonymized responses are found in our replication package~\cite{repl_pack}.

\subsubsection{\textbf{Finding Participants}} Our target population consisted of Mozilla stakeholders with experience in issue resolution.
To identify potential participants, we explored the developers' profiles from Mozilla Research's website~\cite{mozilla_research}, LinkedIn, Mozilla's issue tracker, Mozilla’s Forums~\cite{mozilla_forums}, and Matrix~\cite{mozilla_matrix}.
We created a shortlist of 42 potential participants, all of whom were invited to participate via email.

\subsubsection{\textbf{Participants' Background}} 
Two developers responded to our call and participated in the interview (\ie referred to as D1 and D2). They are current Mozilla developers with 7 to 11 years of experience at the company.
They have extensive issue resolution experience, having resolved around 1.4K issues and contributed to approximately 19K issues in total. %

\subsubsection{\textbf{Response Analysis}} We recorded and transcribed the interviews using Zoom to facilitate response analysis. We corrected inaccuracies in the transcripts, \eg\ misspellings, incorrect phrases, and punctuation. 
Using the revised transcripts, one author analyzed and grouped the participants’ answers to each question into themes representing use cases of the patterns. A second author reviewed the answers and themes for accuracy. Misinterpretations were resolved through discussion.
\looseness=-1
}

\section{Results}
\label{sec:results}

\subsection{\ref{rq:stages}: Issue Resolution Stages}
\label{sub:results_stages}

\rev{\Cref{tab:stages} lists the six identified issue resolution stages and reveals that not all issue reports include all stages, indicating that Firefox developers do not go through these stages, do not need to discuss them in the reports or discuss them in other systems or artifacts (\eg\ instant messaging tools). We discuss the stages starting with the most frequent ones.}
\looseness=-1

\looseness=-1

\subsubsection{\textbf{\impl}} 
This stage is frequently performed and discussed (in 92.1\% of the issues), which is expected as Firefox's issue tracker is integrated with the version control system (Mercurial).
\rev{Among the 28 issues not including any \impl, 25 issues were resolved in other issues, 
two issues were resolved by updating libraries in the host operating system, and the remaining issue was closed after more than four years of being open because the issue was no longer valid.}
\looseness=-1

\subsubsection{\textbf{\crv}} 
This stage is also frequently performed and discussed (in 74.2\% of the issues), which is expected as Firefox's issue tracker is integrated with Firefox's code review tools (\eg Phabricator \cite{phabricator}). 
While \crv is frequently discussed, it is not found in 25.8\% of the reports, especially in defect reports. We found that defects are the least discussed with \crv (31.9\%), compared to enhancement and task reports (6.7\% and 7.7\%). 
\rev{Phabricator~\cite{phabricator}, adopted in 2019, replaced MozReview~\cite{mozreview} and Splinter~\cite{splinter} and became Firefox's only code review tool. Analysis shows that 35\% of defects resolved in or before 2019 lacked \crv, compared to only 17\% after 2019. We found that post-2018 issues without \crv do not include code changes, as the issues were resolved in other issues. %
}

\looseness=-1

\subsubsection{\textbf{\ver}} 

\rev{This stage, covering manual and automatic testing, appears in only 41\% of issues. It is less common in task reports (26.9\%) than in defect and enhancement reports (42.2\% and 41.7\%). Refactoring and testing issues include \ver less often (19.6\% and 31.8\%) compared to implementation-related issues (46.7\%). Categories like Code Improvement, Test Failure, Code Design, and Performance Optimization have few \ver discussions, while Crashes (65\%), Feature Dev. (61.5\%), and UI Issues (63.6\%) show higher inclusion. This indicates still low \ver discussions in the issues and this is consistently found every year.}
\looseness=-1

\subsubsection{\textbf{\ia and \sd}} 

\rev{The \ia and \sd stages are infrequently discussed, appearing in only 37.6\% and 42.1\% of the issues, respectively. Defects are analyzed more frequently (\ie\ \ia in 44.8\% of defects) compared to enhancements (20\%) and tasks (3.9\%), with Crashes, Flaky Tests, Incorrect Page Renderings, and Test Updates being the most analyzed defects. \sd is more common in enhancements (42.2\%) and defects (43.3\%) than in tasks (38.5\%). Refactoring issues are the least analyzed (7.8\%) compared to implementation (42.2\%) and testing-related issues (45.5\%).}
\looseness=-1

\subsubsection{\textbf{\ir}} 
\rev{\ir is the least frequent stage, appearing in only 13.2\% of issues, with just 17\% of defects including it. This suggests that Firefox stakeholders rarely discuss bug reproduction in issue reports.
We observed that \ir is included when bugs are difficult to reproduce or when reproduction is necessary to identify the root cause or localize the bug. The first scenario typically leads to more effort in solving the issues: compared to defects without \ir, defects with \ir take longer to resolve (avg/med: 82.9/5.5 vs. 79.3/18.5 days) and involve more commentators (avg/med: 4.9/4 vs. 7.7/7). 
This is validated by a Mann-Whitney U test~\cite{mcknight2010mann} with $\alpha = 0.05$, with p-values nearly 0.
 The second scenario is supported by the data: 73\% of the issues with \ir include \ia.}

\rqanswer{\textbf{\ref{rq:stages} Findings}: The six stages of issue resolution found in Firefox issue reports appear with varying frequency across different issue reports and problem categories. \impl and \crv are the most frequent stages while \ir is the least frequent.

\looseness=-1}

\subsection{\ref{rq:interactions}: Interactions between Issue Resolution Stages}
\label{sub:results_interactions}
We examined the 356 stage sequences obtained in \Cref{sub:process} by analyzing the frequency of stage bi-grams and tri-grams in the sequences.
Bi-grams are pairs of consecutive stages (\texttt{\textbf{S}},\texttt{\textbf{T}}) in a sequence, which represent possible stage transitions (\texttt{\textbf{S}}\trs\texttt{\textbf{T}}) in the resolution process. In our data, we found all possible bi-grams between stages, except for  \crvs \trs \irs, and five extremely rare transitions (appearing only once or twice): \impls \trs \irs, \vers \trs \irs, \irs \trs \crvs, and \irs \trs \vers. All these transitions include \ir (\irs) as the source or target stage.
\looseness=-1

Of the 1,430 bi-grams found in the sequences, nine are the most recurrent, covering 80.6\% of the bi-gram occurrences: \impls \trs \crvs (403 cases), \crvs \trs \impls (187), \sds \trs \impls (133), \crvs \trs \vers (94), \impls \trs \vers (86), \vers \trs \impls (73), \ias \trs \impls (70), \ias \trs \sds (65), and \crvs \trs \sds~(41). From these transitions, we observe the interplay among \impl, \crv, and \ver, in which \impl undergoes quality assurance activities and these also lead to additional code changes (or to \sd~-- see \crvs \trs \sds above). \sd (\sds) can lead to code changes, and \ia (\ias) can result in code changes or solution design activities. Notably, while \ias \trs \irs and \irs \trs \ias are not among the most frequent bi-grams, they appear in 73.5\% and 70.6\% of the issues containing both stages (34). This further supports our finding that \ia and \ir typically occur together.

The analysis of tri-grams, sets of three consecutive stages in a sequence (\texttt{\textbf{S}}\trs\texttt{\textbf{T}}\trs\texttt{\textbf{U}}), not only provides extra evidence of the interplay among \impl, \crv,  and \ver, but also the relationship among \ia, \sd, \impl, and \crv. Of 1,109 tri-grams found in the sequences, 25 are the most frequent, covering 80.6\% of the tri-grams, with the following three being the most frequent: \impls \trs \crvs \trs \impls (168 occurrences in 97 issues), \crvs \trs \impls \trs \crvs (114 occurrences in 61 issues), and \sds \trs \impls \trs \crvs (85 occurrences in 76 issues). 

\rqanswer{\textbf{\ref{rq:interactions} Findings}: Firefox's developers switch among different resolution stages to solve issues. 
	Stage bi-gram and tri-gram analysis reveal that developers frequently engage in three scenarios: 1) reproducing the issues (\irs) along with issue analysis (\ias)  to confirm the issues and reason about them; 2) analyzing the issues (\ias) along with solution design (\sds), and then engaging in implementation (\impls) and code reviews (\crvs); and 3) inspecting (\crvs) and verifying the implemented solution (\vers), adapting the implementation when needed (\impls).
\looseness=-1}

\subsection{\ref{rq:process}: Issue Resolution Process}
\label{sub:results_process}

\begin{figure}[t]
	\centering
	\includegraphics[width=1\linewidth]{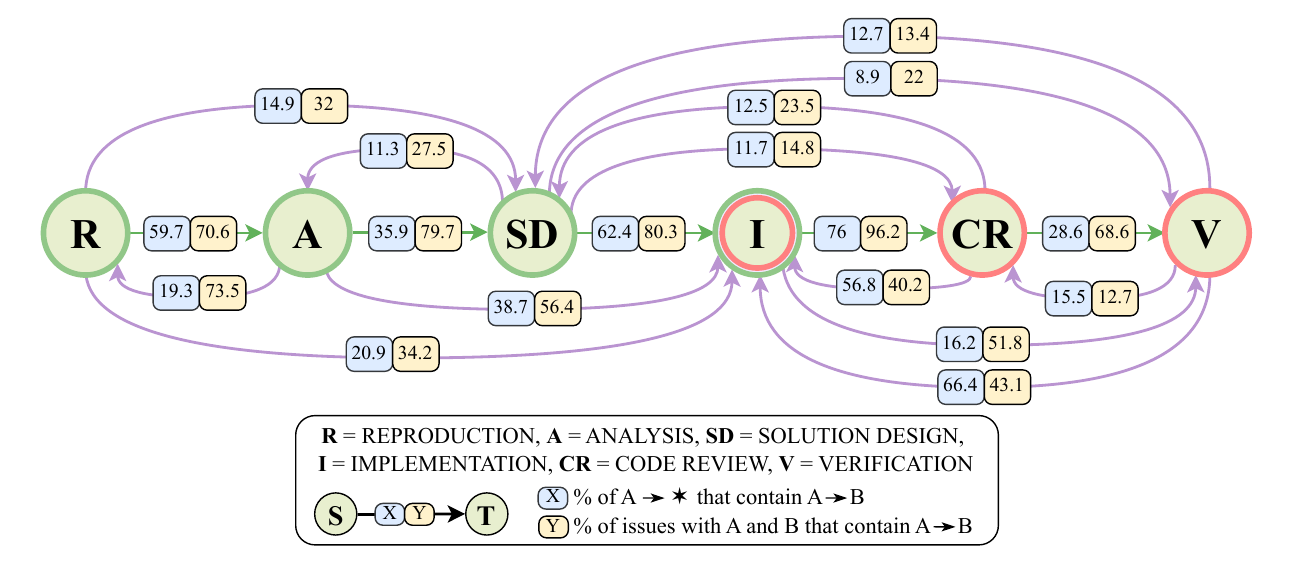}
	\caption{Overall Issue Resolution Process of Firefox}
	\label{fig:overall_process}
\end{figure}

\Cref{fig:overall_process} shows the overall issue resolution process at Firefox, derived from the bi-gram analysis we performed on the 356 stage sequences. The process is a directed graph in which the nodes represent the six resolution stages (\eg~\impl or \impls) and the edges represent transitions between stages (\eg~\impl \trs \crv or \impls \trs \crvs). The nodes with  \colorbox{StateGreen}{green} and \colorbox{StateRed}{red} borders are initial and end nodes, selected from the most frequent initial and end stages in the sequences. All nodes imply a loop to itself, indicating that the stage can be performed multiple times in a row.

The process includes only the most frequent bi-grams \texttt{\textbf{S}}~\trs~\texttt{\textbf{T}}, selected based on the \textit{proportion of bi-grams} starting with \texttt{\textbf{S}} (\texttt{\textbf{S}}\trs \texttt{\textbf{*}}) that contain \texttt{\textbf{S}}\trs \texttt{\textbf{T}}. This proportion is shown in \colorbox{BoxesBlue}{blue boxes} of the edges in \Cref{fig:overall_process}. For example, \impls  \trs \crvs has a frequency of 403/530 =  \colorbox{BoxesBlue}{76\%} since  from all 530 bigrams that start with \impls (\impls \trs \texttt{\textbf{*}}) there are 403 occurrences of \impls  \trs \crvs. The frequencies of all transitions starting from a given node add up to at least 90\%. For example, all transitions coming out of \impls add up to 92.3\%. The \colorbox{BoxesYellow}{yellow boxes} of the edges represent the \textit{proportion of issues} containing \texttt{\textbf{S}} and \texttt{\textbf{T}} that contain the bigram \texttt{\textbf{S}}\trs\texttt{\textbf{T}}. For example, 254 of 264 (\colorbox{BoxesYellow}{96.2\%}) issues with both \impls and \crvs contain \impls \trs \crvs.

We make two observations about the process in \Cref{fig:overall_process}:
\begin{itemize}[itemindent=0cm,leftmargin=0.3cm]
	\item The process deviates from the (theoretical) linear process outlined by the existing literature and Firefox's documentation (see \Cref{sub:background_issue_res,sub:firefox_process}). Instead of a linear sequence of stages (starting with \irs and going through every stage from left to right until \ver and/or \crv are completed --- see the path with green transitions in \Cref{fig:overall_process}), the process is more complicated than expected as it includes iterative interactions between stages. This means developers go back and forth from one stage to another, forming different workflows of issue resolution.
	\item Some nodes have a high number of incoming and outgoing transitions, indicating the level of importance of such stages in the process. Specifically, \sd has four incoming and four outgoing transitions and \impl has five incoming and two outgoing transitions. These stages are pivotal because they allow for stage switches from and to many of the other stages. \crv and \ver are the second most important stages, both having three incoming and three outgoing transitions, while \ia and \ir are less important with fewer transitions.
 \looseness=-1
\end{itemize}

\rqanswer{\textbf{\ref{rq:process} Findings}: Firefox's issue resolution follows an iterative process that deviates from the theoretical linear process. In this process, developers go back and forth from one state to another as needed to solve the issues. \sd and \impl, followed by \crv and \ver, play a key role as they are the source and target for most of the other stages.
}

\subsection{\ref{rq:patterns}: Issue Resolution Patterns}
\label{sub:results_patterns}

\Cref{fig:overall_process} shows that Firefox's issue resolution process differs from the expected linear process from prior work. However, the figure does not show how much the process differs and the different instances of the process that developers follow.

Employing the qualitative approach described in \Cref{sub:resolution_patterns}, we identified 47 distinct instances of the process, which we call \textit{issue resolution patterns}. These patterns appear in the 356 issue reports with varying frequency and complexity: a pattern appears in 1 to 64 reports (7.6 on avg, 4 median), 20 patterns are categorized as complex (they imply high issue resolution effort) and 27 as simple (they imply low resolution effort). Of all patterns, 18 patterns are the most recurrent: they are found in 5 to 64 reports (16 on avg., 9.5 median), covering 287 reports (80.6\%); 12 are simple, and 6 are complex.  \Cref{tab:patterns} shows the 10 most recurrent patterns -- the entire pattern catalog can be found in our replication package~\cite{repl_pack}. 

\subsubsection{\textbf{Pattern Examples}} We describe two patterns of different kinds to illustrate different workflows of issue resolution.

The pattern \texttt{\textbf{A,SD,I,(I$\mid$CR$\mid$V)?}} represents the process in which developers first analyze the reported problem (\ias). They then design the solution (\sds) (\eg propose a potential solution or review a proposed solution), and then implement the solution~(\impls). Developers may then review the code (\crvs) and/or test the code changes to verify if they solved the issue (\vers). Based on QA feedback, more code changes may occur (\impls). This pattern is  \textit{simple} because it includes only three ``mandatory'' stages (\texttt{\textbf{A,SD,I}}) followed by three ``optional'' stages (\texttt{\textbf{(I$\mid$CR$\mid$V)?}}).
\looseness=-1

The pattern {\texttt{\textbf{(SD,I,(CR$\mid$V))+}}} suggests a process in which \sd, \impl, \crv, and/or \ver are performed repetitively to resolve the issue. In the repetitive series, {\texttt{\textbf{(CR$\mid$V)}}} means that either one or both can appear after an \sds and an \impls. To resolve issues, developers perform four distinct stages where all stages are repetitive, making this pattern \textit{complex}.
\looseness=-1

\begin{table}[t]
	\caption{Top 10 Frequent Issue Resolution Patterns}
	\label{tab:patterns}
	\resizebox{\columnwidth}{!}{%
		\begin{tabular}{l|l|c|c}
			\hline
			\multicolumn{1}{c|}{\textbf{\begin{tabular}[c]{@{}c@{}}Pattern\end{tabular}}} & \multicolumn{1}{c|}{\textbf{Description}}                                                                                                                                                                  & \textbf{\begin{tabular}[c]{@{}c@{}}Com-\\ plexity\end{tabular}} & \textbf{\begin{tabular}[c]{@{}c@{}}\# of\\ Issues\end{tabular}} \\ \hline
			\texttt{\textbf{I,CR,I?}}                                                                                            & \begin{tabular}[c]{@{}l@{}}Implement the solution and review the code;\\ followed by another optional implementation.\end{tabular}                                                                         & Simple                                                          & 64                                                           \\ \hline
			\texttt{\textbf{A,I,(I$\mid$CR$\mid$V)?}}                                                                                     & \begin{tabular}[c]{@{}l@{}}Analyze the problem and implement the solution;\\ followed by another optional I or CR or V or\\ any combination.\end{tabular}                                                  & Simple                                                          & 32                                                           \\ \hline
			\texttt{\textbf{(I,(CR$\mid$V))+}}                                                                                        & \begin{tabular}[c]{@{}l@{}}Implement the solution; review the code and/or\\ verify the implementation; I, CR and/or V\\ repeat more than once.\end{tabular}                                                & Complex                                                         & 28                                                           \\ \hline
			\texttt{\textbf{SD,I,CR,(I$\mid$V)?}}                                                                                      & \begin{tabular}[c]{@{}l@{}}Design and implement the solution and review the\\ code; followed by another optional I or V or both.\end{tabular}                                                              & Simple                                                          & 24                                                           \\ \hline
			\texttt{\textbf{A,SD,I,(I$\mid$CR$\mid$V)+}}                                                                                  & \begin{tabular}[c]{@{}l@{}}Analyze the problem, design, and implement the\\ solution; followed by another optional I or CR\\ or V or any combination.\end{tabular}                                         & Simple                                                          & 22                                                           \\ \hline
			\texttt{\textbf{I}}                                                                                                 & Implement the solution.                                                                                                                                                                                    & Simple                                                          & 21                                                           \\ \hline
			\texttt{\textbf{I,CR,V,I?}}                                                                                           & \begin{tabular}[c]{@{}l@{}}Implement the solution, review the code, and verify\\ the implementation; followed by another optional I.\end{tabular}                                                          & Simple                                                          & 16                                                           \\ \hline
			\texttt{\textbf{SD,(I,(CR$\mid$V))+}}                                                                                     & \begin{tabular}[c]{@{}l@{}}Design the solution; implement the solution,\\ review code and/or verify the implementation;\\ I, CR and/or V repeat more than once.\end{tabular}                               & Complex                                                         & 13                                                           \\ \hline
			\texttt{\textbf{(SD,I,(CR$\mid$V))+}}                                                                                      & \begin{tabular}[c]{@{}l@{}}Design and implement the solution; review the\\ code, and/or verify the implementation;\\ SD,I, CR and/or V repeat more than once.\end{tabular}                                 & Complex                                                         & 12                                                           \\ \hline
			\texttt{\textbf{A,(I,(CR$\mid$V))+}}                                                                                     
			& \begin{tabular}[c]{@{}l@{}}Analyze the problem; implement the solution,\\ review code, and/or verify the implementation;\\ I, CR and/or V repeat more than once.\end{tabular}                              & Complex                                                         & 7                                                            
			\\ \hline
			\multicolumn{4}{c}{
				\scriptsize{
					\texttt{\textbf{R}}=\ir, \texttt{\textbf{A}}=\ia, \texttt{\textbf{SD}}=\sd,}} \\
			
			\multicolumn{4}{c}{
				\scriptsize{\texttt{\textbf{I}}=\impl, \texttt{\textbf{CR}}=\crv, \texttt{\textbf{V}}=\ver}}

		\end{tabular}%
	}
\end{table}

\subsubsection{\textbf{Process and Pattern Diversity}} All the identified patterns represent instances of the issue resolution process. The 47 instances indicate a wide variety of ways to solve Firefox issues. While more generalized patterns can be formed from the 47 patterns, our qualitative approach carefully identified the patterns to accurately reflect the observed process from the issues. 
We did not forcefully merge patterns into more general ones but did validate the patterns against the process from \Cref{fig:overall_process} (which was derived quantitatively).
\looseness=-1

The diversity of the patterns/process is observed across problem categories. Six of the 17 problem categories have more unique patterns (14 to 23) than the remaining 11 categories (1 to 11 patterns). These six categories are: Defective Functionality (23 unique patterns found in 43  issues), Code Design (21 patterns in 75  issues), 
UI Issue (22 patterns in  33 issues), Test Failure (17 patterns in 17  issues), Crash (17 patterns in  17 issues), and
Feature Development (14 patterns in 39  issues). 
\looseness=-1

Issue resolution for some categories is more diverse than for other categories, despite having a similar number of issues. For example, UI Issues are solved with 22 patterns, and Code Improvement issues are solved with 11 patterns, despite both categories covering 32-33 issues. We also found that the six most frequent patterns shown in \Cref{tab:patterns} were used to resolve issues of more than half of the categories  (9-11 of 17 categories). This illustrates that the same issue resolution pattern can solve problems of different kinds.
\looseness=-1

The diversity of the patterns/process is also observed throughout Firefox's lifespan, from 2010 to 2023. During these 14 years, the five most frequent patterns (found in 48\% of the issues) were observed in 11 to 14 different years. The 10 most frequent patterns (found in 67\% of the issues) are found in 7 to 14 different years. The 39 patterns appearing in two or more issues were used in 2 to 14 years.

\subsubsection{\textbf{Pattern Complexity and Resolution Effort}} \Cref{tab:patterns_issue_types} shows that \rev{70.8\% of the issues (252 of 356)} are solved with the 27 simple patterns. \rev{These, compared to issues solved with a complex pattern, are solved faster (avg/med: 58/5 vs. 119.8/19.5 days), require fewer stages in the process (avg/med: 2.9/3 vs. 9.9/9), and include fewer commentators (avg/med: 4.4/4 vs. 7.4/7). 
	A Mann–Whitney U test~\cite{mcknight2010mann} (at $\alpha = 0.05$) confirmed these differences across all these factors with \textit{p-value} = 0.0.}
	\looseness=-1

While all the problem kinds are solved with simple patterns in most of the cases (55.8\% -  85.1.7\% of the issues), 5 of 17 categories tend to have more issues solved with complex patterns (104 of 356 = 29.2\%), compared to the other 12 categories. These five categories are:  Code Design (22 of 75 issues are solved with a complex pattern),  Defective Functionality (19 of 43), Feature Development (13 of 39), UI Issue (14 of 33), and Crash (8 of 23). This indicates that these categories contain issues that require more effort to be solved. 
\rev{Despite the issues in these categories being solved with similar resolution time (avg/med: 72.4/9 vs 81.6/6 days), they include more stages in their process (avg/med: 5.2/4 vs. 4/3) and more commentators (avg/med: 5.8/5 vs. 4.4/4) with statistical significance (Mann–Whitney U test, \textit{p-value} = 0.0), suggesting potentially higher resolution effort.}

\looseness=-1

\Cref{tab:patterns_issue_types} also reveals that 76.9\% of the tasks are solved using a simple pattern whereas 70.7\% of the defects and 68.3\% of the enhancements are solved with a simple pattern. \rev{As pattern complexity suggests, compared to defects and enhancements, tasks require less effort: they are solved significantly faster (avg/med: 9/4 vs. 82.3/7 and 76.9/11.5), require fewer process stages  (avg/med: 3.6/2 vs. 5.2/4 and 5/4), and include fewer commentators (avg/med: 4/3 vs. 5.4/4 and 5.3/5). These results are statistically significant, according to the Mann–Whitney U test (p-values = 0.02, 0.03, and 0.002, respectively.)}
\looseness=-1

\rqanswer{\textbf{\ref{rq:patterns} Findings}: The 47 identified issue resolution patterns indicate that solving issues at Firefox is done in a wide variety of ways. Of these, 18 patterns are recurrently found in 80.6\% of Firefox issue reports. 
The process of issue resolution in Firefox is diverse and far from linear. The diverse and iterative nature of the process is consistently observed throughout Firefox's 14 years of evolution.}

\begin{table}[t]
\caption{Number of Issues Across Issue Types}
\label{tab:patterns_issue_types}
\centering
\begin{tabular}{c|cc|c}
	\hline
	\multirow{2}{*}{\textbf{Issue Type}} & \multicolumn{2}{c|}{\textbf{Pattern Complexity}}        & \multirow{2}{*}{\textbf{Total}} \\ \cline{2-3}
	& \multicolumn{1}{c|}{\textbf{Complex}} & \textbf{Simple} &                                 \\ \hline
	\textbf{Defect}                      & \multicolumn{1}{c|}{79}               & 191             & \textbf{270}                    \\ \hline
	\textbf{Enhancement}                 & \multicolumn{1}{c|}{19}               & 41              & \textbf{60}                     \\ \hline
	\textbf{Task}                        & \multicolumn{1}{c|}{6}                & 20              & \textbf{26}                     \\ \hline
	\textbf{Total}                       & \multicolumn{1}{c|}{\textbf{104}}     & \textbf{252}    & \textbf{356}                    \\ \hline
\end{tabular}
\end{table}

\rev{
\subsection{\ref{rq:pattern_usefulness}: Use Cases for the Issue Resolution Patterns}
\label{sub:pattern_usefulness}

The two interviewed Mozilla developers (\ie D1 and D2) identified the following use cases for the derived patterns:

\subsubsection{\textbf{Identifying Issues with a Complex Resolution}} 
D1 and D2 suggested that the patterns could help detect issues with complex resolutions, especially those involving repetitive stages, which may signal excessive time and effort spent by developers. D1 noted that such patterns could indicate when a bug takes an unexpectedly complex path, explaining, ``If you went through three different implementations and three different verifications and it still didn't work, something went wrong here." D2 also emphasized the value of a tool that identifies these complex issues, saying, ``It might be interesting to have some sort of tool that watches the bugs and when it sees this snowball effect... it could alert a product person that this bug is chewing up a lot of time." Both developers agreed on the importance of early detection of complex resolutions. Both emphasized that detecting these issues would help understand why a process is taking longer than expected, allowing for timely corrections. 
	
\subsubsection{\textbf{Identifying Issues Not Following the Expected Process}} 
D1 suggested that a tool could be useful for detecting issues that deviate from expected workflows, especially those requiring human verification. He explained, ``Some bugs require human verification... it requires installing third-party software on a machine. [...] And it's up to developers to highlight when this is the case." D1 emphasized that a tool could help by alerting developers when an issue seems to need third-party support for verification, stating, ``If you had a tool that said, hey, this bug that you open looks like it might need some third-party support for verification, that might actually be a helpful thing."

\subsubsection{\textbf{Identifying Potentially Complex Code Components}} 
D2 suggested that tracking the complexity of issue resolution in specific Mozilla Firefox code components or modules could reveal underlying quality issues, such as technical debt or accumulated code complexity. As D2 put it, ``You could track the complexity of the issue resolution process in a given module... and get insights like, hey, it looks like most of the time when you touch this area of code it ends up being a slog." D2 emphasized that such insights could signal the need for refactoring, stating, ``Maybe it's time to refactor this? Maybe it's time to clean this up... this part of the code base is a tar pit and we probably want to spend some resources making it less ornery." 
\looseness=-1

\subsubsection{\textbf{Improving Bots to Detect Unsolvable Issues}} 
D2 suggested that the resolution patterns could help improve the heuristics of existing bots (\eg bugbug~\cite{bugbug}) used to process Mozilla issues, enabling them to better identify issues that are unsolvable or particularly challenging to solve. As D2 explained, ``If you could identify signs that this bug is not going to be solved or is about to fall through the cracks, that would be pretty cool." He noted that current bots already attempt to detect when issues have ``fallen through the cracks" and need attention, saying, ``We have some bots that do that kind of work." 
\looseness=-1
	
\subsubsection{\textbf{Suggesting and Decomposing Meta-Issues}} 
Both D1 and D2 suggested that issues with complex resolutions, as indicated by the complex patterns, might represent meta-issues: large issues that could be broken down into smaller, more manageable issues. They proposed that a tool capable of flagging these cases and suggesting possible decompositions would be highly beneficial. As D1 explained, ``We have this idea of a meta bug, which is a bug which hosts a whole bunch of related bugs," and suggested the tool could flag such cases and offer ideas like, "could this be split? Here are some topics that it sounds like you could split this down into."
	
\subsubsection{\textbf{Training Junior Developers}} 
D1 and D2 both indicated that the patterns could serve as valuable training tools for junior Mozilla developers, offering insights into the practical aspects of issue resolution. D1 noted that junior developers often have high expectations and approach issues linearly, seeking a perfect solution. However, D1 emphasized that the patterns show the issue-resolution process is typically incremental and iterative, involving multiple cycles of code review and verification. As D1 explained, ``You will probably have to iterate... you will probably have to go through this solution more than once. And that's okay. That's expected. It's part of the job." 

Both developers suggested that automation is needed to realize such use cases, particularly tools that identify patterns in issue discussions and classify them as simple or complex. Our future work will develop such tools to automate the identification of textual content in issue comments (\eg issue resolution activities) and use algorithms to derive sequences of stages and patterns. This process will likely combine machine learning with heuristic-based approaches.

}

\rqanswer{\textbf{\ref{rq:pattern_usefulness} Findings}: \rev{The interviewed Mozilla developers suggested that the resolution patterns could help identify complex issues, workflow deviations, and low-quality code components, improve bots for detecting unsolvable issues, decompose large issues, and train junior developers. 
}
}

\section{Discussion and Implications}
\label{sec:implication_usefulness}

\textbf{Firefox's Issue Resolution in Practice.} 
Our study highlights the iterative and diverse nature of Firefox's issue resolution process, which widely deviates from the theoretical linear models often assumed in the literature (\eg Rajlich's incremental change process~\cite{rajlich2011software}). Instead of following a straightforward path, developers address various types of issues by moving back and forth through multiple stages as needed. This reflects the iterative and incremental approach characteristic of modern software development, aligning more closely with agile methodologies than with rigid frameworks like Waterfall~\cite{rajlich2011software}.

\rev{
	\textbf{Patterns Generalizability.} 
	While the results only apply to Firefox, we conducted a small case study that annotated 20 issue reports (of different kinds)
    from two open-source projects: Chromium~\cite{chromium} and GnuCash~\cite{gnucash}. The goal was to validate if these projects follow resolution patterns similar to Firefox's.
	Details of the study methodology are found in our replication package~\cite{repl_pack}.
	
	We identified seven distinct resolution patterns for the 10 Chromium issue reports, all of which correspond to Firefox's patterns. Three patterns, \texttt{\textbf{`I,CR'}}, \texttt{\textbf{`I,CR,V'}}, and \texttt{\textbf{`SD,I,CR,V'}} appeared in two issues each, aligning with Firefox patterns \texttt{\textbf{`I,CR,I?'}}, \texttt{\textbf{`I,CR,V,I?'}}, and \texttt{\textbf{`SD,I,CR,(I$\mid$V)'}} which appeared in 64, 16, and 24 issues respectively.
	Notably, these Firefox patterns are among the top seven most recurrent patterns, which strengthens pattern generalizability (they are found in Chromium issues).
	As for GnuCash, we identified 10 resolution patterns across the 10 issues, with nine of these patterns aligning with nine of the 47 Firefox patterns. The nine Firefox patterns are fairly common as they were observed in 4 to 22 issues. The GnuCash pattern \texttt{\textbf{`(A,I)+'}} does not have any corresponding Firefox pattern.
	
	The 7 and 10 patterns identified for Chromium and GnuCash indicate that developers in these projects also employ diverse approaches to issue resolution. This suggests that some Firefox patterns may generalize across projects of varying scales and governance. However, a large-scale study with a statistically significant sample is needed to confirm these observations.

}

\section{Threats to Validity}
\label{sec:threats}

\textbf{Construct and Internal Validity}. 
\rev{Relying solely on issue reports poses a validity threat. Issue discussions may not capture all of Firefox's resolution activities, either because certain actions do not require documentation or were discussed/recorded in other artifacts or channels. This limitation may explain why some stages (\eg issue reproduction) are absent in certain issues. Consequently, the derived patterns should be interpreted with caution, as they reflect the resolution process \textit{documented in issue reports}, which may differ from the practical process. However, according to Firefox's documentation~\cite{firefox-patching}, issue reports are one of the primary artifacts for tracking Firefox changes, and developers are encouraged to document relevant problem information within them. Moreover, we are confident in the accuracy of traces for implementation, code review, and verification stages, due to the tool integration with the issue tracker, as well as the requirement for verification to mark an issue as ``VERIFIED." This provides confidence that issue report discussions capture the implemented resolution process.}

Researcher subjectivity and potential confirmation bias introduced during issue coding, resolution pattern inference, and results interpretation represent key validity threats.
To address these,  we implemented a rigorous open-coding methodology involving multiple coding phases.
Each issue report was reviewed multiple times, accompanied by discussion sessions between annotators. Both annotators critically annotated and verified the data at each phase, resolving disagreements through consensus. The results interpretation was thoroughly discussed and supported by data-driven evidence.

\textbf{External Validity}.  Our pattern catalog and results may not generalize to all issues from Firefox and to other systems, as is typical in case studies. 
This stems from the relatively small set of issues we coded to derive the patterns. 
To strengthen generalization, our study analyzed a statistically significant sample, in which the distribution of coded issues resembles that of all Firefox issues. \rev{We also 
annotated 20 issue reports of Chromium and GnuCash, and found that some of the most frequent Firefox patterns cover the resolution workflows found in the 20 issues, which implies that at least some of the derived patterns can be generalized to these projects. While the results are indicative, in-depth studies are needed to confirm these results and establish generalizability.}

\section{Related Work}
\label{sec:related_work}

Researchers have proposed a variety of techniques to address issue management challenges and automate several tasks in the process~\cite{zou2018practitioners,Adnan:msr25}. For example, researchers have proposed automated techniques to better report issues~\cite{song_toward_2022,Fazzini:TSE22,song2023burt}, assess issue quality~\cite{mahmud:icpc2025,chaparro2019assessing,chaparro2017detecting,song2020bee}, 
predict the priority and severity of the issues~\cite{umer2019cnn,tian2015automated}, categorize issue types~\cite{somasundaram2012automatic,catolino2019not}, assign developers to issues~\cite{xia2013accurate,chaitra2022bug}, suggest potential duplicate issues~\cite{Zhou2012a,he2020duplicate,zhang2023duplicate}, reproduce buggy behavior~\cite{Zhao2019,feng2022gifdroid,zhang2023automatically,Zhao2019,zhao2022recdroid+}, localize buggy code files~\cite{akbar2020large,lee2018bench4bl,Ye2016b,ciborowska2022fast,Wong2014,Kochhar2014,florez2021combining,chaparro2019using,chaparro2017using,chaparro2019reformulating,saha2024toward,mahmud2024using,chaparro2016reduction},  and predict re-opened issues~\cite{zimmermann2012characterizing,shihab2010predicting}. 

Researchers have studied issues for a variety of purposes: to understand  decision-making~\cite{hesse2016documented} and 
the discourse used to describe issues~\cite{chaparro2017detecting,chaparro2016vocabulary}; extract decision information~\cite{mahadi2020cross}; understand stakeholders' information needs~\cite{Breu2010}; characterize/predict different kinds of issues such as \textit{won't fix} issues~\cite{panichella2021won}, fixed/resolved issues~\cite{Guo2010}, non-reproducible bugs~\cite{rahman2022works}, and bug/issue types~\cite{Tan2014,catolino2019not,limsettho2016unsupervised}; predict issue severity~\cite{Sureka2010}; understand workarounds~\cite{yan2023programmers} and visual content in issues~\cite{agrawal2022understanding};  
questions~\cite{huang2019empirical}, and information types in issues~\cite{arya2019analysis}.

Researchers have used automated mining of issue data (\eg status changes) and version control/code review data to identify development processes and assess delays and inconsistencies~\cite{marques2018assessing,krismayer2019using}. 
They have utilized process mining techniques to integrate data from different sources (\eg VCS, issue trackers, and mail archives)~\cite{poncin2011process, gupta2014process,mittal2014process} and proposed process mining techniques~\cite{rubin2007process,gupta2014nirikshan,saini2020control} to gain insight into development processes. Other work has studied the life cycle of issues by mining and analyzing issue state transitions~\cite{eren2023analyzing,dobrzynski2016tracing,wang2012predicting,coremans2023process}.
These works focused more broadly on issue management and identified transitions of issue states (\eg from ``Assigned" to ``In progress" to ``Closed"). 
However, issue states are often too broad to provide detailed insights into how stakeholders resolve issues in practice.

Unlike prior work, our research qualitatively analyzed issue reports to identify resolution stages, develop a process model, and uncover detailed patterns of issue resolution at Firefox. This in-depth analysis led to new insights into the issue resolution process. To our knowledge, we are the first to examine how the issue resolution process is actually implemented and discussed in practice, and how it differs from the theoretical models found in the literature.

\section{Conclusions}
\label{sec:conclusions}
We conducted a case study to understand the process employed by Mozilla Firefox developers to solve software issues. By implementing a multi-coder open-coding methodology, we qualitatively analyzed the issue report comments, identified six issue resolution stages, and derived an overall process model. We found 47 issue resolution patterns, which are instances of the process and represent how Firefox developers resolve issues in practice. This process is iterative and deviates widely from the theoretical linear process from the literature.

\section*{acknowledgements}
We thank Trevor Stalnaker, Nathan Wintersgill, Nadeeshan De Silva, Mehedi Sun, and Md Akram Khan for assisting with issue report annotation. This work is supported by U.S. NSF grant CCF-2239107. The opinions, findings, and conclusions expressed in this paper are those of the authors and do not necessarily reflect the sponsors' opinions.
\looseness=-1

\balance
\bibliographystyle{IEEEtran}
\bibliography{references}

\begin{thebibliography}{100}
\providecommand{\url}[1]{#1}
\csname url@samestyle\endcsname
\providecommand{\newblock}{\relax}
\providecommand{\bibinfo}[2]{#2}
\providecommand{\BIBentrySTDinterwordspacing}{\spaceskip=0pt\relax}
\providecommand{\BIBentryALTinterwordstretchfactor}{4}
\providecommand{\BIBentryALTinterwordspacing}{\spaceskip=\fontdimen2\font plus
\BIBentryALTinterwordstretchfactor\fontdimen3\font minus
  \fontdimen4\font\relax}
\providecommand{\BIBforeignlanguage}[2]{{%
\expandafter\ifx\csname l@#1\endcsname\relax
\typeout{** WARNING: IEEEtran.bst: No hyphenation pattern has been}%
\typeout{** loaded for the language `#1'. Using the pattern for}%
\typeout{** the default language instead.}%
\else
\language=\csname l@#1\endcsname
\fi
#2}}
\providecommand{\BIBdecl}{\relax}
\BIBdecl

\bibitem{Zimmermann2009}
T.~Zimmermann, R.~Premraj, J.~Sillito, and S.~Breu, ``Improving bug tracking
  systems,'' in \emph{ICSE'09}, 2009, pp. 247--250.

\bibitem{Zimmermann2010}
T.~Zimmermann, R.~Premraj, N.~Bettenburg, S.~Just, A.~Schr\"{o}ter, and
  C.~Weiss, ``What {Makes} a {Good} {Bug} {Report}?'' \emph{TSE}, vol.~36,
  no.~5, pp. 618--643, 2010.

\bibitem{firefox-bug-handling}
``Firefox's bug handling documentation,''
  \url{https://firefox-source-docs.mozilla.org/bug-mgmt/index.html}, 2024.

\bibitem{zhang2016literature}
T.~Zhang, H.~Jiang, X.~Luo, and A.~T. Chan, ``A literature review of research
  in bug resolution: Tasks, challenges and future directions,'' \emph{The
  Computer Journal}, vol.~59, no.~5, pp. 741--773, 2016.

\bibitem{saha2015understanding}
R.~K. Saha, S.~Khurshid, and D.~E. Perry, ``Understanding the triaging and
  fixing processes of long lived bugs,'' \emph{Information and software
  technology}, vol.~65, pp. 114--128, 2015.

\bibitem{zeller2009programs}
A.~Zeller, \emph{Why programs fail: a guide to systematic debugging}.\hskip 1em
  plus 0.5em minus 0.4em\relax Elsevier, 2009.

\bibitem{rajlich2011software}
V.~Rajlich, \emph{Software engineering: The current practice}.\hskip 1em plus
  0.5em minus 0.4em\relax Crc Press, 2011.

\bibitem{eren2023analyzing}
{\c{C}}.~Eren, K.~{\c{S}}ahin, and E.~T{\"u}z{\"u}n, ``Analyzing bug life
  cycles to derive practical insights,'' in \emph{EASE'23}, 2023, pp. 162--171.

\bibitem{repl_pack}
``Online replication package,'' \url{https://doi.org/10.5281/zenodo.14727541},
  2024.

\bibitem{bugzillaBugzilla}
``Bugzilla,'' \url{https://www.bugzilla.org/}, 2024.

\bibitem{jira}
``Jira,'' \url{https://www.atlassian.com/software/jira}, 2024.

\bibitem{github}
``Github,'' \url{https://github.com/features/issues}, 2024.

\bibitem{bertram2009social}
D.~Bertram, ``The social nature of issue tracking in software engineering,''
  \emph{University of Calgary}, 2009.

\bibitem{bertram2010communication}
D.~Bertram, A.~Voida, S.~Greenberg, and R.~Walker, ``Communication,
  collaboration, and bugs: the social nature of issue tracking in small,
  collocated teams,'' in \emph{CSCW'10}, 2010, pp. 291--300.

\bibitem{xia2013accurate}
X.~Xia, D.~Lo, X.~Wang, and B.~Zhou, ``Accurate developer recommendation for
  bug resolution,'' in \emph{WCRE'13}, 2013, pp. 72--81.

\bibitem{bugzila-bug-lifecycle}
``The life cycle of a bug in bugzilla,''
  \url{https://www.bugzilla.org/docs/2.18/html/lifecycle.html}, 2024.

\bibitem{mozilla-firefox}
``Firefox browsers,'' \url{https://www.mozilla.org/en-US/firefox/}, 2024.

\bibitem{firefox-bug-1029919}
``Firefox bug \#1029919,''
  \url{https://bugzilla.mozilla.org/show_bug.cgi?id=1029919}, 2024.

\bibitem{firefox-bug-1718748}
``Firefox bug \#1718748,''
  \url{https://bugzilla.mozilla.org/show_bug.cgi?id=1718748}, 2024.

\bibitem{mozregression}
``mozregression,'' \url{https://mozilla.github.io/mozregression/}, 2024.

\bibitem{phabricator}
``Phabricator,'' \url{https://phabricator.services.mozilla.com/}, 2024.

\bibitem{firefox-patching}
``Firefox: how to submit a patch,''
  \url{https://firefox-source-docs.mozilla.org/contributing/how_to_submit_a_patch.htm},
  2024.

\bibitem{firefox-code-quality}
``Firefox: Code quality,''
  \url{https://firefox-source-docs.mozilla.org/code-quality/index.html}, 2024.

\bibitem{firefox-reviewer-checklist}
``Firefox: Reviewer checklist,''
  \url{https://firefox-source-docs.mozilla.org/contributing/reviewer_checklist.html},
  2024.

\bibitem{firefox-security-bugs}
``Firefox: Fixing security bugs,''
  \url{https://firefox-source-docs.mozilla.org/bug-mgmt/processes/fixing-security-bugs.html},
  2024.

\bibitem{Working-Firefox}
``Working on firefox,''
  \url{https://firefox-source-docs.mozilla.org/contributing/index.html}, 2024.

\bibitem{firefox-bug-pipeline}
``Firefox's bug pipeline documentation,'' \url{https://tinyurl.com/2up6wjp3},
  2024.

\bibitem{mozilla-bug-types}
``Mozilla's bug types,''
  \url{https://firefox-source-docs.mozilla.org/bug-mgmt/guides/bug-types.html},
  2024.

\bibitem{firefox-triage}
``Firefox's bug triage,''
  \url{https://firefox-source-docs.mozilla.org/bug-mgmt/policies/triage-bugzilla.html},
  2024.

\bibitem{firefox-feature-triage}
``Firefox's new feature triage,''
  \url{https://firefox-source-docs.mozilla.org/bug-mgmt/policies/new-feature-triage.html},
  2024.

\bibitem{firefox-security-approval}
``Firefox: Approving security bugs,''
  \url{https://firefox-source-docs.mozilla.org/bug-mgmt/processes/security-approval.html},
  2024.

\bibitem{firefox-try}
``Pushing to try,''
  \url{https://firefox-source-docs.mozilla.org/tools/try/index.html}, 2024.

\bibitem{testing-firefox-ml}
A.~Halberstadt and M.~Castelluccio, ``Testing firefox more efficiently with
  machine learning,''
  \url{https://hacks.mozilla.org/2020/07/testing-firefox-more-efficiently-with-machine-learning/},
  2020.

\bibitem{shipping-firefox}
``Pocket guide: Shipping firefox,''
  \url{https://firefox-source-docs.mozilla.org/contributing/pocket-guide-shipping-firefox.html},
  2024.

\bibitem{mozilla-products}
``Mozilla products in bmo,''
  \url{https://bugzilla.mozilla.org/describecomponents.cgi}, 2024.

\bibitem{bugzilla-api}
``Bugzilla's rest api,'' \url{https://wiki.mozilla.org/Bugzilla:REST_API},
  2024.

\bibitem{spencer2009card}
D.~Spencer, \emph{Card sorting: Designing usable categories}.\hskip 1em plus
  0.5em minus 0.4em\relax Rosenfeld Media, 2009.

\bibitem{hypothesis}
``The hypothesis web annotation tool,'' \url{https://web.hypothes.is}, 2024.

\bibitem{Cohen}
J.~Cohen, ``A coefficient of agreement for nominal scales,'' \emph{Educational
  and psychological measurement}, vol.~20, no.~1, pp. 37--46, 1960.

\bibitem{firefox-bug}
``Firefox bug \#1363344,'' \url{https://tinyurl.com/mr3d7h6e}, 2023.

\bibitem{firefox-bug-991812}
``Firefox bug \#991812,''
  \url{https://bugzilla.mozilla.org/show_bug.cgi?id=991812}, 2024.

\bibitem{firefox-bug-698552}
``Firefox bug \#698552,''
  \url{https://bugzilla.mozilla.org/show_bug.cgi?id=698552}, 2024.

\bibitem{firefox-bug-676248}
``Firefox bug \#676248,''
  \url{https://bugzilla.mozilla.org/show_bug.cgi?id=676248}, 2024.

\bibitem{firefox-bug-730907}
``Firefox bug \#730907,''
  \url{https://bugzilla.mozilla.org/show_bug.cgi?id=730907}, 2024.

\bibitem{mozilla_research}
``\url{https://research.mozilla.org/},'' 2024.

\bibitem{mozilla_forums}
``\url{https://www.mozilla.org/en-US/about/forums/},'' 2024.

\bibitem{mozilla_matrix}
``\url{https://wiki.mozilla.org/Matrix},'' 2024.

\bibitem{mozreview}
``Mozreview,''
  \url{https://wiki.mozilla.org/EngineeringProductivity/Projects/MozReview},
  2024.

\bibitem{splinter}
``Splinter,'' \url{https://wiki.mozilla.org/BMO/Splinter}, 2024.

\bibitem{mcknight2010mann}
P.~E. McKnight and J.~Najab, ``Mann-whitney u test,'' \emph{The Corsini
  encyclopedia of psychology}, pp. 1--1, 2010.

\bibitem{bugbug}
``\url{https://github.com/mozilla/bugbug},'' 2024.

\bibitem{chromium}
``Chromium,'' \url{https://www.chromium.org/Home/}, 2024.

\bibitem{gnucash}
``Gnucash,'' \url{https://www.gnucash.org/}, 2024.

\bibitem{zou2018practitioners}
W.~Zou, D.~Lo, Z.~Chen, X.~Xia, Y.~Feng, and B.~Xu, ``How practitioners
  perceive automated bug report management techniques,'' \emph{TSE}, vol.~46,
  no.~8, pp. 836--862, 2018.

\bibitem{Adnan:msr25}
A.~Adnan, A.~Saha, and O.~Chaparro, ``Sprint: An assistant for issue report
  management,'' in \emph{MSR'25}, 2025.

\bibitem{song_toward_2022}
Y.~Song, J.~Mahmud, Y.~Zhou, O.~Chaparro, K.~Moran, A.~Marcus, and
  D.~Poshyvanyk, ``Toward interactive bug reporting for ({Android} app)
  end-users,'' in \emph{FSE'22}, 2022.

\bibitem{Fazzini:TSE22}
M.~Fazzini, K.~P. Moran, C.~Bernal-Cardenas, T.~Wendland, A.~Orso, and
  D.~Poshyvanyk, ``Enhancing mobile app bug reporting via real-time
  understanding of reproduction steps,'' \emph{TSE}, 2022.

\bibitem{song2023burt}
Y.~Song, J.~Mahmud, N.~De~Silva, Y.~Zhou, O.~Chaparro, K.~Moran, A.~Marcus, and
  D.~Poshyvanyk, ``Burt: A chatbot for interactive bug reporting,'' in
  \emph{ICSE'23}, 2023.

\bibitem{mahmud:icpc2025}
J.~Mahmud, A.~Saha, O.~Chaparro, K.~Moran, and A.~Marcus, ``Combining language
  and app ui analysis for the automated assessment of bug reproduction steps,''
  in \emph{ICPC'25}, 2025.

\bibitem{chaparro2019assessing}
O.~Chaparro, C.~Bernal-C{\'a}rdenas, J.~Lu, K.~Moran, A.~Marcus, M.~Di~Penta,
  D.~Poshyvanyk, and V.~Ng, ``Assessing the quality of the steps to reproduce
  in bug reports,'' in \emph{ESEC/FSE'19}, 2019.

\bibitem{chaparro2017detecting}
O.~Chaparro, J.~Lu, F.~Zampetti, L.~Moreno, M.~Di~Penta, A.~Marcus, G.~Bavota,
  and V.~Ng, ``Detecting missing information in bug descriptions,'' in
  \emph{FSE'17}, 2017.

\bibitem{song2020bee}
Y.~Song and O.~Chaparro, ``Bee: A tool for structuring and analyzing bug
  reports,'' in \emph{ESEC/FSE'20}, 2020.

\bibitem{umer2019cnn}
Q.~Umer, H.~Liu, and I.~Illahi, ``Cnn-based automatic prioritization of bug
  reports,'' \emph{IEEE Transactions on Reliability}, vol.~69, no.~4, pp.
  1341--1354, 2019.

\bibitem{tian2015automated}
Y.~Tian, D.~Lo, X.~Xia, and C.~Sun, ``Automated prediction of bug report
  priority using multi-factor analysis,'' \emph{ESE}, vol.~20, pp. 1354--1383,
  2015.

\bibitem{somasundaram2012automatic}
K.~Somasundaram and G.~C. Murphy, ``Automatic categorization of bug reports
  using latent dirichlet allocation,'' in \emph{ISEC'12}, 2012, pp. 125--130.

\bibitem{catolino2019not}
G.~Catolino, F.~Palomba, A.~Zaidman, and F.~Ferrucci, ``Not all bugs are the
  same: Understanding, characterizing, and classifying bug types,'' \emph{JSS},
  vol. 152, pp. 165--181, 2019.

\bibitem{chaitra2022bug}
B.~Chaitra and K.~Swarnalatha, ``Bug triaging: right developer recommendation
  for bug resolution using data mining technique,'' in \emph{ERCICA'22}.\hskip
  1em plus 0.5em minus 0.4em\relax Springer, 2022, pp. 609--618.

\bibitem{Zhou2012a}
J.~Zhou and H.~Zhang, ``Learning to rank duplicate bug reports,'' in
  \emph{CIKM'12}, 2012, pp. 852--861.

\bibitem{he2020duplicate}
J.~He, L.~Xu, M.~Yan, X.~Xia, and Y.~Lei, ``Duplicate bug report detection
  using dual-channel convolutional neural networks,'' in \emph{ICPC'20}, 2020,
  pp. 117--127.

\bibitem{zhang2023duplicate}
T.~Zhang, D.~Han, V.~Vinayakarao, I.~C. Irsan, B.~Xu, F.~Thung, D.~Lo, and
  L.~Jiang, ``Duplicate bug report detection: How far are we?'' \emph{TOSEM},
  vol.~32, no.~4, pp. 1--32, 2023.

\bibitem{Zhao2019}
Y.~Zhao, T.~Yu, T.~Su, Y.~Liu, W.~Zheng, J.~Zhang, and W.~G. Halfond,
  ``Recdroid: Automatically reproducing android application crashes from bug
  reports,'' in \emph{ICSE'19}, 2019, pp. 128--139.

\bibitem{feng2022gifdroid}
S.~Feng and C.~Chen, ``Gifdroid: an automated light-weight tool for replaying
  visual bug reports,'' in \emph{ICSE'22}, 2022.

\bibitem{zhang2023automatically}
Z.~Zhang, R.~Winn, Y.~Zhao, T.~Yu, and W.~G. Halfond, ``Automatically
  reproducing android bug reports using natural language processing and
  reinforcement learning,'' in \emph{ISSTA'23}, 2023, pp. 411--422.

\bibitem{zhao2022recdroid+}
Y.~Zhao, T.~Su, Y.~Liu, W.~Zheng, X.~Wu, R.~Kavuluru, W.~G. Halfond, and T.~Yu,
  ``Recdroid+: Automated end-to-end crash reproduction from bug reports for
  android apps,'' \emph{TOSEM}, vol.~31, no.~3, pp. 1--33, 2022.

\bibitem{akbar2020large}
S.~A. Akbar and A.~C. Kak, ``A large-scale comparative evaluation of ir-based
  tools for bug localization,'' in \emph{MSR'20}, 2020, pp. 21--31.

\bibitem{lee2018bench4bl}
J.~Lee, D.~Kim, T.~F. Bissyand{\'e}, W.~Jung, and Y.~Le~Traon, ``Bench4bl:
  reproducibility study on the performance of ir-based bug localization,'' in
  \emph{ISSTA'18}, 2018, pp. 61--72.

\bibitem{Ye2016b}
X.~Ye, H.~Shen, X.~Ma, R.~Bunescu, and C.~Liu, ``From word embeddings to
  document similarities for improved information retrieval in software
  engineering,'' in \emph{ICSE'16}, 2016, pp. 404--415.

\bibitem{ciborowska2022fast}
A.~Ciborowska and K.~Damevski, ``Fast changeset-based bug localization with
  bert,'' in \emph{ICSE'22}, 2022, pp. 946--957.

\bibitem{Wong2014}
C.-P. Wong, Y.~Xiong, H.~Zhang, D.~Hao, L.~Zhang, and H.~Mei, ``Boosting
  bug-report-oriented fault localization with segmentation and stack-trace
  analysis,'' in \emph{ICSME'14}, 2014, pp. 181--190.

\bibitem{Kochhar2014}
P.~S. Kochhar, Y.~Tian, and D.~Lo, ``Potential biases in bug localization: Do
  they matter?'' in \emph{ASE'14}, 2014, pp. 803--814.

\bibitem{florez2021combining}
J.~M. Florez, O.~Chaparro, C.~Treude, and A.~Marcus, ``Combining query
  reduction and expansion for text-retrieval-based bug localization,'' in
  \emph{SANER'21}, 2021, pp. 166--176.

\bibitem{chaparro2019using}
O.~Chaparro, J.~M. Florez, and A.~Marcus, ``Using bug descriptions to
  reformulate queries during text-retrieval-based bug localization,''
  \emph{EMSE}, vol.~24, pp. 2947--3007, 2019.

\bibitem{chaparro2017using}
------, ``Using observed behavior to reformulate queries during text
  retrieval-based bug localization,'' in \emph{ICSME'17}, 2017, pp. 376--387.

\bibitem{chaparro2019reformulating}
O.~Chaparro, J.~M. Florez, U.~Singh, and A.~Marcus, ``Reformulating queries for
  duplicate bug report detection,'' in \emph{SANER'19}, 2019, pp. 218--229.

\bibitem{saha2024toward}
A.~Saha, Y.~Song, J.~Mahmud, Y.~Zhou, K.~Moran, and O.~Chaparro, ``Toward the
  automated localization of buggy mobile app uis from bug descriptions,'' in
  \emph{ISSTA'24}, 2024, pp. 1249--1261.

\bibitem{mahmud2024using}
J.~Mahmud, N.~De~Silva, S.~A. Khan, S.~H. Mostafavi, S.~H. Mansur, O.~Chaparro,
  A.~Marcus, and K.~Moran, ``On using gui interaction data to improve text
  retrieval-based bug localization,'' in \emph{ICSE'24}, 2024.

\bibitem{chaparro2016reduction}
O.~Chaparro and A.~Marcus, ``On the reduction of verbose queries in text
  retrieval based software maintenance,'' in \emph{ICSE'16}, 2016.

\bibitem{zimmermann2012characterizing}
T.~Zimmermann, N.~Nagappan, P.~J. Guo, and B.~Murphy, ``Characterizing and
  predicting which bugs get reopened,'' in \emph{ICSE'12}, 2012, pp.
  1074--1083.

\bibitem{shihab2010predicting}
E.~Shihab, A.~Ihara, Y.~Kamei, W.~M. Ibrahim, M.~Ohira, B.~Adams, A.~E. Hassan,
  and K.-i. Matsumoto, ``Predicting re-opened bugs: A case study on the eclipse
  project,'' in \emph{WCRE'10}, 2010, pp. 249--258.

\bibitem{hesse2016documented}
T.-M. Hesse, V.~Lerche, M.~Seiler, K.~Knoess, and B.~Paech, ``Documented
  decision-making strategies and decision knowledge in open source projects: An
  empirical study on firefox issue reports,'' \emph{IST}, vol.~79, pp. 36--51,
  2016.

\bibitem{chaparro2016vocabulary}
O.~Chaparro, J.~M. Florez, and A.~Marcus, ``On the vocabulary agreement in
  software issue descriptions,'' in \emph{ICSME'16}, 2016.

\bibitem{mahadi2020cross}
A.~Mahadi, K.~Tongay, and N.~A. Ernst, ``Cross-dataset design discussion
  mining,'' in \emph{SANER'20}, 2020, pp. 149--160.

\bibitem{Breu2010}
S.~Breu, R.~Premraj, J.~Sillito, and T.~Zimmermann, ``Information {Needs} in
  {Bug} {Reports}: {Improving} {Cooperation} {Between} {Developers} and
  {Users},'' in \emph{CSCW'10}, 2010, pp. 301--310.

\bibitem{panichella2021won}
S.~Panichella, G.~Canfora, and A.~Di~Sorbo, ``“won't we fix this issue?”
  qualitative characterization and automated identification of wontfix issues
  on github,'' \emph{IST}, vol. 139, p. 106665, 2021.

\bibitem{Guo2010}
P.~J. Guo, T.~Zimmermann, N.~Nagappan, and B.~Murphy, ``Characterizing and
  predicting which bugs get fixed: An empirical study of {Microsoft}
  {Windows},'' in \emph{ICSE'10}, 2010, pp. 495--504.

\bibitem{rahman2022works}
M.~M. Rahman, F.~Khomh, and M.~Castelluccio, ``Works for me! cannot
  reproduce--a large scale empirical study of non-reproducible bugs,''
  \emph{EMSE}, vol.~27, no.~5, p. 111, 2022.

\bibitem{Tan2014}
L.~Tan, C.~Liu, Z.~Li, X.~Wang, Y.~Zhou, and C.~Zhai,
  ``\BIBforeignlanguage{en}{Bug characteristics in open source software},''
  \emph{\BIBforeignlanguage{en}{EMSE}}, vol.~19, no.~6, pp. 1665--1705, 2014.

\bibitem{limsettho2016unsupervised}
N.~Limsettho, H.~Hata, A.~Monden, and K.~Matsumoto, ``Unsupervised bug report
  categorization using clustering and labeling algorithm,'' \emph{JSEKE'16},
  vol.~26, no.~07, pp. 1027--1053, 2016.

\bibitem{Sureka2010}
A.~Sureka and P.~Jalote, ``Detecting {Duplicate} {Bug} {Report} {Using}
  {Character} {N}-{Gram}-{Based} {Features},'' in \emph{ASPEC'10}, 2010, pp.
  366--374.

\bibitem{yan2023programmers}
A.~Yan, H.~Zhong, D.~Song, and L.~Jia, ``How do programmers fix bugs as
  workarounds? an empirical study on apache projects,'' \emph{EMSE}, vol.~28,
  no.~4, p.~96, 2023.

\bibitem{agrawal2022understanding}
V.~Agrawal, Y.-H. Lin, and J.~Cheng, ``Understanding the characteristics of
  visual contents in open source issue discussions: a case study of jupyter
  notebook,'' in \emph{EASE'22}, 2022, pp. 249--254.

\bibitem{huang2019empirical}
Y.~Huang, D.~A. da~Costa, F.~Zhang, and Y.~Zou, ``An empirical study on the
  issue reports with questions raised during the issue resolving process,''
  \emph{EMSE}, vol.~24, pp. 718--750, 2019.

\bibitem{arya2019analysis}
D.~Arya, W.~Wang, J.~L. Guo, and J.~Cheng, ``Analysis and detection of
  information types of open source software issue discussions,'' in
  \emph{ICSE'19}, 2019, pp. 454--464.

\bibitem{marques2018assessing}
R.~Marques, M.~M. da~Silva, and D.~R. Ferreira, ``Assessing agile software
  development processes with process mining: A case study,'' in \emph{CBI'18},
  vol.~1, 2018, pp. 109--118.

\bibitem{krismayer2019using}
T.~Krismayer, C.~Mayr-Dorn, J.~Tuder, R.~Rabiser, and P.~Gr{\"u}nbacher,
  ``Using constraint mining to analyze software development processes,'' in
  \emph{ICSSP'19}, 2019, pp. 94--103.

\bibitem{poncin2011process}
W.~Poncin, A.~Serebrenik, and M.~Van Den~Brand, ``Process mining software
  repositories,'' in \emph{CSMR'11}, 2011, pp. 5--14.

\bibitem{gupta2014process}
M.~Gupta, A.~Sureka, and S.~Padmanabhuni, ``Process mining multiple
  repositories for software defect resolution from control and organizational
  perspective,'' in \emph{MSR'14}, 2014, pp. 122--131.

\bibitem{mittal2014process}
M.~Mittal and A.~Sureka, ``Process mining software repositories from student
  projects in an undergraduate software engineering course,'' in
  \emph{ICSE'14}, 2014, pp. 344--353.

\bibitem{rubin2007process}
V.~Rubin, C.~W. G{\"u}nther, W.~M. Van Der~Aalst, E.~Kindler, B.~F. Van~Dongen,
  and W.~Sch{\"a}fer, ``Process mining framework for software processes,'' in
  \emph{ICSP'07}, 2007, pp. 169--181.

\bibitem{gupta2014nirikshan}
M.~Gupta and A.~Sureka, ``Nirikshan: Mining bug report history for discovering
  process maps, inefficiencies and inconsistencies,'' in \emph{ISEC'14}, 2014,
  pp. 1--10.

\bibitem{saini2020control}
V.~Saini, P.~Singh, and A.~Sureka, ``Control-flow based anomaly detection in
  the bug-fixing process of open-source projects,'' in \emph{ISEC'20}, 2020,
  pp. 1--11.

\bibitem{dobrzynski2016tracing}
B.~Dobrzy{\'n}ski and J.~Sosnowski, ``Tracing life cycle of software bugs,'' in
  \emph{DepCoS-RELCOMEX'16}, 2016, pp. 109--120.

\bibitem{wang2012predicting}
J.~Wang and H.~Zhang, ``Predicting defect numbers based on defect state
  transition models,'' in \emph{ESEM'12}, 2012, pp. 191--200.

\bibitem{coremans2023process}
B.~Coremans, A.~L. Klomp, S.~A. Rukmono, J.~Kr{\"u}ger, D.~Fahland, and M.~R.
  Chaudron, ``Process mining from jira issues at a large company,'' in
  \emph{ICSME'23}, 2023, pp. 425--435.

\end{thebibliography}

\end{document}